\documentclass[prl,aps,twocolumn,footinbib,superscriptaddress,notitlepage,floatfix]{revtex4-2}
\usepackage{amsmath}
\usepackage{amssymb}
\usepackage{graphicx}
\usepackage{dcolumn}
\usepackage{enumerate}
\usepackage{bm}
\usepackage[normalem]{ulem}
\usepackage{xcolor}
\usepackage{comment}
\usepackage{subfigure}
\usepackage{breqn}
\usepackage{hyperref}
\usepackage{braket}
\usepackage[sectionbib]{bibunits}
\defaultbibliographystyle{apsrev4-2}
\defaultbibliography{library}
\hypersetup{colorlinks=true, citecolor=blue, urlcolor=blue, linkcolor=blue}
\usepackage{environ}
\NewEnviron{equationS}{%
\begin{align}\begin{split}
  \BODY
\end{split}\end{align}
}

\newcommand{\Rev}[1]{#1} 
\newcommand{\ghost}[1]{{\color{red}{}\normalcolor}}

\makeatletter
\let\cat@comma@active\@empty
\newcommand*{\balancecolsandclearpage}{%
  \close@column@grid
  \clearpage
  \twocolumngrid
}
\makeatother

\begin{document}
\title{Trade-offs between unitary and measurement induced spin squeezing in cavity QED}
\author{Diego Barberena}
\email{diego.barberena@colorado.edu}
\affiliation{JILA, NIST and Department of Physics, University of Colorado, Boulder, Colorado 80309, USA}
\affiliation{Center for Theory of Quantum Matter, University of Colorado, Boulder, Colorado 80309, USA}
\author{Anjun Chu}
\affiliation{JILA, NIST and Department of Physics, University of Colorado, Boulder, Colorado 80309, USA}
\affiliation{Center for Theory of Quantum Matter, University of Colorado, Boulder, Colorado 80309, USA}
\author{James K. Thompson}
\affiliation{JILA, NIST and Department of Physics, University of Colorado, Boulder, Colorado 80309, USA}
\author{Ana Maria Rey}
\affiliation{JILA, NIST and Department of Physics, University of Colorado, Boulder, Colorado 80309, USA}
\affiliation{Center for Theory of Quantum Matter, University of Colorado, Boulder, Colorado 80309, USA}
\date{\today}

\begin{abstract}
We study the combined effects of measurements and unitary evolution on the preparation of spin squeezing in an ensemble of atoms interacting with a single electromagnetic field mode inside a cavity. We derive simple criteria that determine the conditions at which measurement based entanglement generation overperforms unitary protocols. We include all relevant sources of decoherence and study both their effect on the optimal spin squeezing and the overall size of the measurement noise, which limits the dynamical range of quantum-enhanced phase measurements. Our conclusions are relevant for state-of-the-art atomic clocks that aim to operate below the standard quantum limit.  
\end{abstract}

\maketitle
\begin{bibunit}
{\it Introduction:} Within the field of quantum metrology~\cite{Degen2017,Pezze2018}, spin squeezed states~\cite{Kitagawa-Ueda1993,Ma2011} constitute a concrete example of a quantum-enhanced resource with near-term practical applications. Their ability to measure spin rotations with a sensitivity that surpasses the standard quantum limit (SQL), i.e. the fundamental limit on phase estimation achievable with $N$ uncorrelated particles, provides the opportunity for practical metrological gain in e.g. atomic clocks~\cite{Pedrozo-Penafiel2020,Robinson2022}, magnetometers~\cite{Budker2007,Thiele2018,Zheng2020} and matter-wave interferometers~\cite{Kasevich1991,Cronin2009,Greve2022}. Consequently, schemes for efficient spin squeezing preparation~\cite{Kuzmich1998,Kitagawa-Ueda1993,QinChenWangMiranowiczNori+2020+4853+4868}, and experimental demonstrations in a variety of quantum platforms~\cite{Muessel2014,Bohnet2016,Schleier-Smith2010c,Leroux2010,Cox2016,Hosten2016,Braverman2019,Pedrozo-Penafiel2020} have attracted considerable attention.

Particularly promising strategies for the scalable generation of squeezing are provided by QED cavities, where a shared light field mediates all-to-all interactions among atoms inside of a cavity. When driven by an external laser, the resulting dispersive atomic response is nonlinear and can be interpreted as an infinite range unitary Ising interaction called one-axis-twising (OAT)~\cite{Kitagawa-Ueda1993}. This is known to create spin squeezing~\cite{Kitagawa-Ueda1993} and this specific drive-induced mechanism is known as cavity-feedback squeezing~\cite{Leroux2010,Zhang2015,Braverman2019,Pedrozo-Penafiel2020}. On the other hand, after atoms and light interact, photons leaking out of the cavity carry information about the atomic ensemble~\cite{Schleier-Smith2010c,Cox2016,Hosten2016} that can be accessed by continuously monitoring the output light via quantum non-demolition (QND) measurements~\cite{Braginsky1980,Braginsky1996,Kuzmich1998}. Adequate use of this information allows for the estimation of the number of non-excited atoms, which decreases the noise of the state along the magnetization axis, leading to spin squeezing.

In this paper we examine the possible advantages of combining both methods of preparation. We employ a general analytical framework, analogous but distinct to the one presented in Ref.~\cite{Li2022}, that considers the effects of finite detection efficiency, and include from the outset fundamental sources of noise and dissipation.

Our main result can be stated succinctly: when the detection efficiency of the QND measurement is above 0.19, QND outperforms OAT. Otherwise, the choice between QND or OAT depends on other experimentally relevant parameters such as spin flip probability, cavity cooperativity, and atom number [see Fig.~\ref{fig:NoSpinFlip}(c) for details]. We also perform a systematic study of the area of the generated measurement noise, which negatively impacts the dynamical range and utility of the state for quantum-enhanced phase measurements~\cite{Braverman2018}.

\begin{figure}
    \centering
    \includegraphics[width=0.48\textwidth]{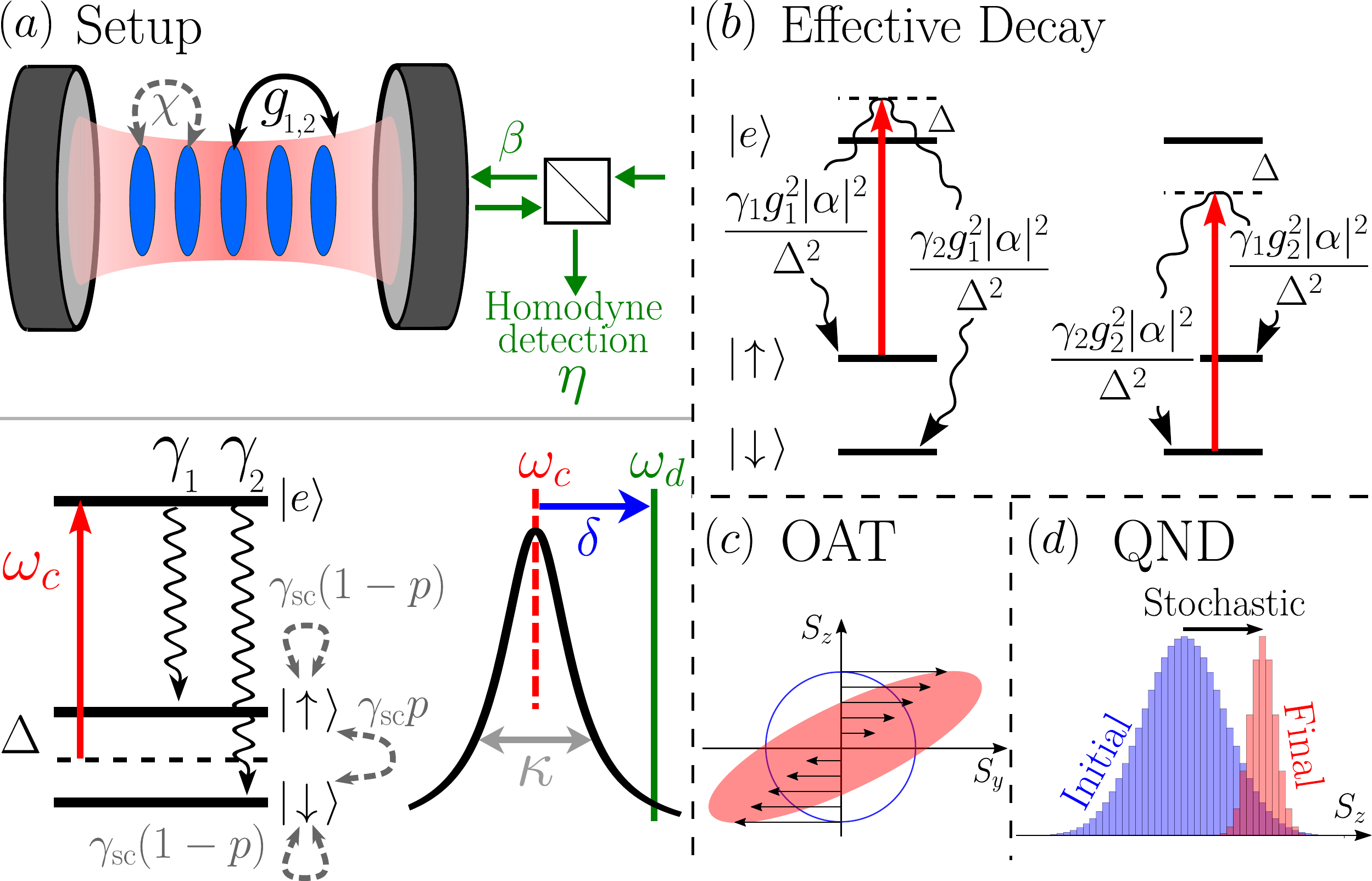}
    \caption{(a) Schematic of the model: a three level system interacting with a QED cavity, which is in turn driven by a laser. Output light is measured via homodyne detection with efficiency $\eta$. Dashed gray lines represent effective processes ($\chi$, $\gamma_{\text{sc}}$). (b) Effective single particle processes in ground manifold. (c) OAT dynamics shears the noise distribution, causing it to get squeezed. (d) Schematic of QND, showing pre (blue) and post (red) measurement distribution in the basis of $\hat{S}_z$.}
    \label{fig:Schematic}
\end{figure}

{\it Model:} We begin with a simple model that exemplifies the physics that we are trying to describe~\cite{Davis2016} [see Fig.~\ref{fig:Schematic}(a)]. We consider an ensemble of $N$ atoms with three-levels: $\ket{\uparrow}$, $\ket{\downarrow}$ and $\ket{e}$ in the level configuration shown in Fig.~\ref{fig:Schematic}(a). The excited state has a finite lifetime, and decays to $\ket{\uparrow}$ and $\ket{\downarrow}$ with rates $\gamma_1$ and $\gamma_2$, respectively. The atoms interact with a single mode of a single port QED cavity, with resonance frequency $\omega_c$, which is detuned from resonance with the $\ket{\uparrow},\ket{\downarrow}\to\ket{e}$ transitions by $\pm \Delta$, as illustrated in Fig.~\ref{fig:Schematic}(a). The cavity is in turn driven close to resonance by a laser tone at frequency $\omega_d$ [detuning $\delta=\omega_d-\omega_c$, see Fig.~\ref{fig:Schematic}(a)] and input flux of $|\beta|^2$ photons per second, and.the transitions $\ket{\uparrow},\ket{\downarrow}\leftrightarrow \ket{e}$ are coupled to the cavity with single photon Rabi frequencies $2g_1$ and $2g_2$, respectively. The light that comes out of the system can then be measured in a homodyne configuration with detection efficiency $\eta$.

Under conditions (to be stated later) that permit adiabatic elimination of the excited state $\ket{e}$ and the cavity degree of freedom, the system evolves under an effective Ito stochastic differential equation~\cite{SM,wiseman_milburn_2009}

\begin{align}\begin{split}\label{eqn:EffDyn}
    d\hat{\rho}=&\bigg\{-i\chi\Big[\hat{S}_z^2,\hat{\rho}\Big]+\Gamma\mathcal{L}_{\hat{S}_z}(\hat{\rho}) \\
    &+\frac{\gamma_{sc}}{2}\sum_{k=1}^N\frac{(1-p)}{2}\mathcal{L}_{\hat{\sigma}_{z}^k}(\hat{\rho})+p\Big[ \mathcal{L}_{\hat{\sigma}_{+}^k}(\hat{\rho})+\mathcal{L}_{\hat{\sigma}_{-}^k}(\hat{\rho})\Big]\bigg\}\,dt\\[5pt]
    &+\sqrt{\Gamma\eta}\Big(\hat{S}_z\hat{\rho}+\hat{\rho}\hat{S}_z-2\braket{\hat{S}_z}\hat{\rho}\Big)\,dW,
\end{split}\end{align}
where $\hat{S}_{x,y,z}=\sum_{k=1}^N\hat{\sigma}_{x,y,z}^k/2$ are collective spin operators acting on the ground manifold $\ket{\uparrow},\ket{\downarrow}$, $\hat{\sigma}_{x,y,z}^k$ are Pauli matrices acting on atom $k$, and $\chi,\Gamma,\gamma_{\text{sc}},p$ are effective parameters related to experimental quantities (we will state their precise definition further on). Unitary dynamics is described by the parameter $\chi$. Incoherent evolution is expressed in terms of Lindbladians $\mathcal{L}_{\hat{L}}(\hat{\rho})\equiv \hat{L}^\dagger\hat{\rho}\hat{L}-\{\hat{L}^\dagger\hat{L},\hat{\rho}\}/2$, and includes collective dephasing ($\Gamma$) and single particle  spin conserving [$\gamma_{\text{sc}}(1-p)$] and spin changing ($\gamma_{\text{sc}}p$) incoherent processes. The final line of Eq.~(\ref{eqn:EffDyn}) incorporates continuous measurement of $\hat{S}_z$ via homodyne detection (in an appropriately chosen quadrature)~\cite{wiseman_milburn_2009,Jacobs2006} with efficiency $\eta$ and includes a stochastic Wiener increment $dW$ to model the probabilistic nature of quantum measurements. The output of the measurements is a time-dependent current $i(t)\equiv dq/dt=2\sqrt{\Gamma\eta}\braket{\hat{S}_z}+dW/dt$. 

To unpack the content of Eq.~(\ref{eqn:EffDyn}) in detail, we consider an initial condition where all atoms begin in the superposition $\ket{\uparrow}+\ket{\downarrow}$, which is relevant to experimental implementations and corresponds to a Bloch vector entirely polarized along the $x$ direction, i.e. $\braket{\hat{S}_x}=N/2$. Furthermore, to obtain a manageable set of equations we use a large $N$ approximation, in which the state remains gaussian, but relax these assumptions later. In this limit  the Bloch vector remains polarized along $x$ but relaxes due to $\gamma_{\text{sc}}$ according to $\hat{S}_x=Ne^{-\gamma_{\text{sc}} t/2}/2$. Fluctuations perpendicular to the Bloch vector satisfy~\cite{Madsen2004,zhang2023stochastic,SM}

\begin{align}\label{eqn:LargeNFluc}
    \begin{split}
        \dot{v}_{zz}&=-\Gamma\eta N v_{zz}^2-2\gamma_{\text{sc}}p(v_{zz}-1)\\
        \dot{v}_{yy}&=2\chi N v_{zy}  e^{-\gamma_{\text{sc}} t/2}+\Gamma N e^{-\gamma_{\text{sc}} t}\\
        &\hspace{2.82cm}-\Gamma N\eta v_{zy}^2-\gamma_{\text{sc}}(v_{yy}-1)\\
        \dot{v}_{zy}&=\chi N e^{-\gamma_{\text{sc}}\frac{t}{2}}v_{zz}-\Gamma \eta N v_{zz} v_{zy}-\frac{\gamma_{\text{sc}}}{2}(2p+1) v_{zy}, 
    \end{split}
\end{align}
where $v_{ab}=\big(2\braket{\{\hat{S}_a,\hat{S}_b\}}-4\braket{\hat{S}_a}\braket{\hat{S}_b}\big)/N$ (for $a,b=z,y$) are (co)variances normalized to the spin projection noise. The equation for $v_{zz}$ ($\propto\hat{S}_z$ variance) evolves under two competing effects: measurements ($\Gamma N\eta$) reveal information about the magnetization and thus reduce $v_{zz}$ [see Fig.~\ref{fig:Schematic}(d)]. On the other hand, spin flips ($\gamma_{\text{sc}}p$) restore the variance to its initial uncorrelated  value $v_{zz}=1$. For $v_{yy}$ ($\hat{S}_y$ variance), single particle processes ($\gamma_{\text{sc}}$) also restore the variance  to its uncorrelated value, but measurement backaction ($\Gamma N$) instead increases the variance. Furthermore, coherent interactions ($\chi N$) mix $v_{yy}$ with $v_{zy}$ and leave $v_{zz}$ untouched, reflecting the shearing dynamics characteristic of OAT [see Fig.~\ref{fig:Schematic}(c)] that leads to a noise distribution squeezed along an intermediate direction in the $\hat{S}_z/\hat{S}_y$ plane. Note that $dW$ does not appear in these equations, indicating that the dynamics they describe does not depend on the specific measurement outcomes.

Measurements do introduce small stochastic corrections to the orientation of the Bloch vector that manifest as deflections in the $yz$ plane. In a small time interval $dt$, these deflections satisfy~\cite{SM}
\begin{align}\label{eqn:LargeNMean}
    \begin{split}
        dz&=-\gamma_{\text{sc}}p\,z\,dt+\sqrt{\Gamma\eta N} v_{zz}\,dW\\
        dy&=(\chi Ne^{-\gamma_{\text{sc}}t/2}z-\gamma_{\text{sc}} y/2)\,dt+\sqrt{\Gamma \eta N} v_{zy}\, dW,
    \end{split}
\end{align}
where $z=\braket{\hat{S}_z}/\sqrt{N/4}$ and $y=\braket{\hat{S}_y}/\sqrt{N/4}$. The measured current evolves according to $dq=\sqrt{\Gamma N\eta}\,z\,dt+dW$ and is connected to $y$ and $z$ through the common increment $dW$. To take advantage of the measurement process these deflections need to be calculated accurately using $i(t)$, since they are different for each measurement realization. Neglecting this information leads to an average state that is not squeezed in any directions. 

The absolute scale for time is set by $\gamma_{\text{sc}}$, which is the total scattering rate from the excited state induced by the probe [see Fig.~\ref{fig:Schematic}(b)]:
\begin{equation}
    \gamma_{\text{sc}}=\frac{\gamma_1g_1^2|\alpha|^2}{\Delta^2}+\frac{\gamma_1g_2^2|\alpha|^2}{\Delta^2}+\frac{\gamma_2g_1^2|\alpha|^2}{\Delta^2}+\frac{\gamma_2g_2^2|\alpha|^2}{\Delta^2},
\end{equation}
where $|\alpha|^2=\kappa|\beta|^2/(\delta_*^2+\kappa^2/4)$ is the number of circulating photons in the cavity, found by multiplying the incident photon number, $|\beta|^2$ [See Fig.~1(a)], by the cavity buildup factor $\kappa/(\delta_*^2+\kappa^2/4)$, and $\delta_*=\delta-(g_1^2-g_2^2)N/2\Delta$ is the detuning of the drive with respect to the dressed cavity mode~(see SM~\cite{SM}). The rest of effective parameters can be expressed in terms of $\gamma_{\text{sc}}$, $d=2\delta_*/\kappa$ and $C=4g_1^2/(\kappa\gamma_1)=4g_2^2/(\kappa\gamma_2)$, the single particle cooperativity, which is a property of cavity geometry:

\begin{equation}\label{eqn:EffParam}
    \chi=\frac{C\gamma_{\text{sc}}d/2}{1+d^2},\hspace{0.3cm}\Gamma=\frac{C\gamma_{\text{sc}}}{1+d^2},\hspace{0.3cm}p=\frac{2\gamma_1\gamma_2}{(\gamma_1+\gamma_2)^2}\leq\frac{1}{2}
\end{equation}
The spin flip probability $p$ measures the relative importance of single particle spin changing processes relative to spin conserving processes, both of which arise through virtual excitation of the excited state and subsequent decay into the ground manifold [see Fig.~\ref{fig:Schematic}(b)].

Cavity feedback squeezing arises when $\delta_*\gg \kappa$. Then $\chi\gg \Gamma$ and OAT dominates until single particle processes disrupt the generation of spin squeezing. QND measurements operate in the opposite regime, with $\delta_*=\chi=0$ and $\Gamma$ maximal. In the absence of $\gamma_{\text{sc}}$, the resulting evolution continuously projects the system onto an $\hat{S}_z$ eigenstate, reducing the variance of $\hat{S}_z$ even beyond the gaussian limit [see Fig.~\ref{fig:Schematic}(d)]. However, the precise eigenstate onto which the system is projected is stochastic and must be estimated accurately using the measurement record.

Adiabatic elimination gives rise to Eq.~(\ref{eqn:EffDyn}) when $\Delta\gg \gamma_{1,2},2g_{1,2}|\alpha|,2g_{1,2}\sqrt{N}$. These conditions guarantee that the excited state is never appreciably populated and that the atom-cavity interaction is dispersive.

{\it Squeezing and state area:} Equations~(\ref{eqn:LargeNFluc}) must be solved with initial conditions $v_{zz}(0)=v_{yy}(0)=1$ and $v_{zy}(0)=0$. Within the gaussian regime, the evolution generates a noise distribution on  the $yz$ plane in the form of a  ellipse whose axes have minimum (maximum) length $v_{\text{min}}$ ($v_{\text{max}}$)~\cite{SM}, and in terms of which we define
\begin{equation}\label{eqn:optsqueeze}
    \xi^2=e^{\gamma_{\text{sc}}t }v_{\text{min}},\hspace{0.5cm} A=e^{\gamma_{\text{sc}}t}\sqrt{v_{\text{min}}v_{\text{max}}}.
\end{equation}
The Wineland squeezing parameter $\xi^2$~\cite{Bollinger1992} quantifies the metrological enhancement of phase measurements compared to uncorrelated atoms (SQL) and includes the effects of reduced contrast. The state area $A$ measures the size of the noise distribution, normalized to the length of the Bloch vector squared.  Under ideal evolution, ($\gamma_{\text{sc}}= 0$ and  $\eta= 1$), $A$ remains  of order  1, but loss of information  leads to an area that can be substantially larger. An increase in $A$
reduces the metrological utility of the generated squeezing  since it limits the range of  phases that can be measured with some degree of quantum enhancement~\cite{Braverman2018}. 

\begin{figure}
    \centering
    \includegraphics[width=0.48\textwidth]{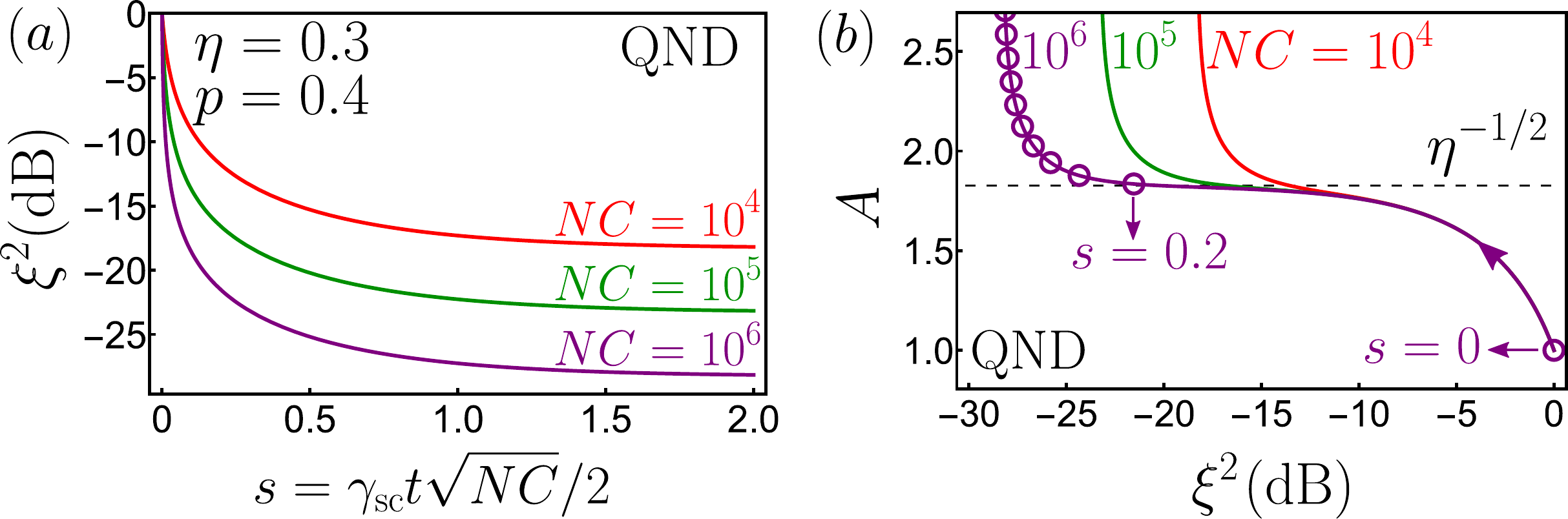}
    \caption{(a) Squeezing ($\xi^2$) as a function of time for various $NC$ in the QND configuration. (b) QND squeezing vs. state area ($A$) plotted parametrically with time as a parameter. Circles are equally spaced in $s$ intervals of size $0.2$.}
    \label{fig:Measurement}
\end{figure}

{\it Measurement limit:} Here $\delta_*=\chi=0$ and $\Gamma=C\gamma_{\text{sc}}$. Assuming that $NC\eta\gg 1$, simple analytic solutions can be written for the fluctuations and the estimator of $\braket{\hat{S}_z}$~\cite{SM}. The minimum-variance axis lies along $z$, giving rise to a Wineland parameter of 
\begin{equation}\label{eqn:MeasurementOptimalSqueezing}
    \xi^2_{t}=v_{zz}=\sqrt{\frac{2p}{NC\eta}},
\end{equation} within the timescale $\tau=(N C\eta p)^{-1/2}/(2\gamma_{\text{sc}})$, while $v_{yy}$ grows as $1+NC\gamma_{\text{sc}} t$ and $v_{zy}=0$. The subscript $t$ in $\xi^2_{t}$ indicates that $\xi^2$ has been optimized over time. $\xi^2$ is depicted as a function of $s=\gamma_{\text{sc}}t\sqrt{NC}/2$ for different values of $NC$ in Fig.~\ref{fig:Measurement}(a).

 Waiting for a few $\tau$ times gets $\xi^2$ closer to $\xi^2_t$, but waiting for too long leads to uncontrolled growth of $v_{yy}$ and hence of state area. We show this in Fig.~\ref{fig:Measurement}(b), where we plot $A$ vs. $\xi^2$ parametrically as a function of time. The sharp upward turn in the curve indicates that $A$ is growing without any improvement in $\xi^2$. Notice also that there are plateaus of constant $A=\eta^{-1/2}$, more visible at larger $NC$. In these plateaus spin flips are not yet active, so the decrease in $\hat{S}_z$ variance is exactly compensated by the increase in $\hat{S}_y$ variance.

{\it Unitary limit:} Here $\eta=0$ and $\delta_*\gg \kappa$. All the relevant equations are now linear and can be solved exactly, but we consider the effects of $\gamma_{\text{sc}}p$ on $\xi^2$ perturbatively. Assuming $\chi N t\gg 1$, this leads to~\cite{SM,Chu2021}
\begin{equation}\label{eqn:UnitSqueezing}
    \xi^2\approx \frac{1}{\chi^2N^2 t^2}+\frac{\Gamma/\chi}{\chi N t}+\frac{2}{3}\gamma_{\text{sc}} p t.
\end{equation}
The first and second terms include the effects of interactions ($\chi$) and collective dephasing ($\Gamma$), respectively. The third term is due to spin flips and is the main obstruction for unitary  spin squeezing. 

\begin{figure}
    \centering
    \includegraphics[width=0.48\textwidth]{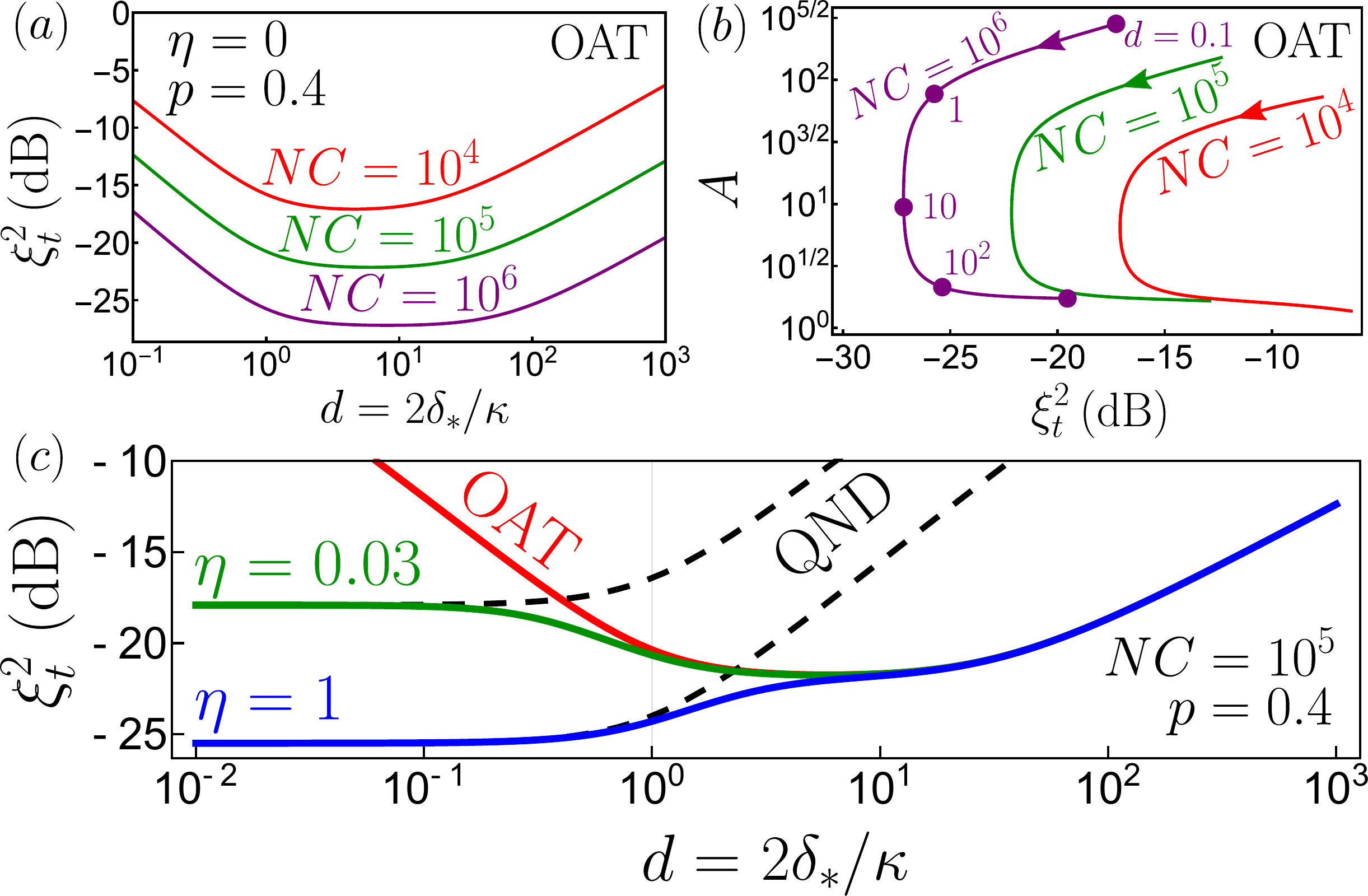}
    \caption{(a) Time optimized squeezing ($\xi^2_t$) as a function of detuning in the OAT configuration for various $NC$. (b) OAT squeezing vs. state area plotted parametrically with $d$ as a parameter. Each filled circle occurs at a value of $d$ 10 times bigger than the previous one. (c) Squeezing optimized over time as a function of $d$ at fixed $p=0.4$ and $NC=10^5$ for different $\eta$. Solid red is OAT ($\eta=0$) and dashed black is pure QND ($\hat{S}_z$ variance).}
    \label{fig:Unitary}
\end{figure}

\begin{figure}
    \centering
    \includegraphics[width=0.48\textwidth]{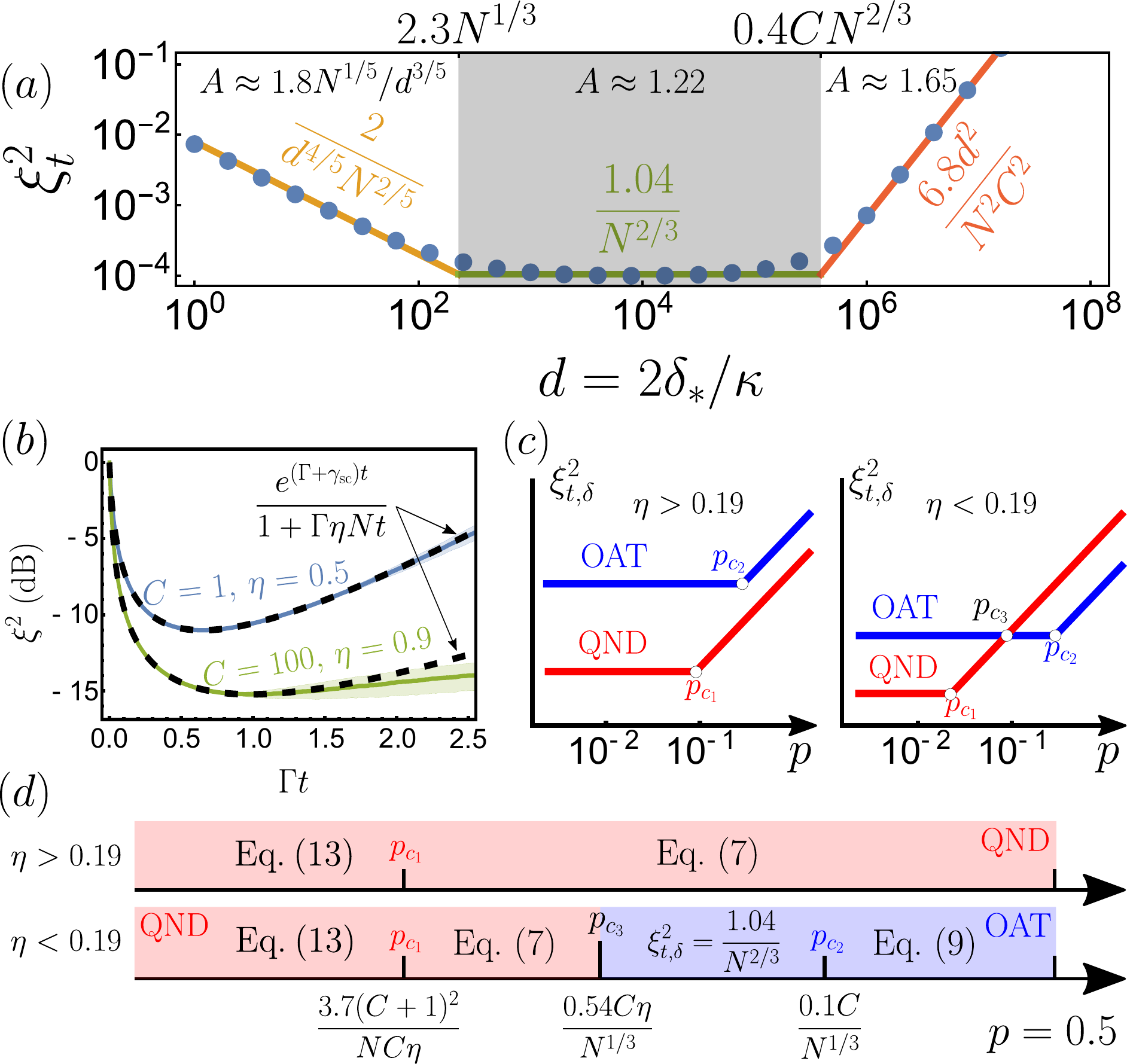}
    \caption{(a) Time optimized spin squeezing $\xi^2_t$ as a function of $d$ when $p=0$, $\eta=0$, $C=10^2$ and $N=10^6$. (b) Time profile of  $\xi^2$ for two values of $(\eta,C)$ averaged over measurement realizations. Shaded areas indicate the dispersion of $\xi^2$ values over different individual measurements. Dashed black is an analytical model, with $\Gamma=C\gamma_{\text{sc}}$. (c) Optimal spin squeezing for OAT and QND as a function of $p$ for different $\eta$. (d) Summary of results.}
    \label{fig:NoSpinFlip}
\end{figure}

The behaviour of $\xi^2_t$ (time optimized squeezing) with $d$ depends on whether collective dephasing ($\Gamma$) is active at the optimal squeezing time or not. If $\Gamma$ is active, relevant for smaller values of $d$, then $\xi^2_t$ arises from the competition between the second and third terms in Eq.~(\ref{eqn:UnitSqueezing}), leading to $\xi^2_t=\xi^2_{t,\delta}\sqrt{1+1/d^{2}}$, where
\begin{align}\begin{split}\label{eqn:OptimalUnitarySqueezing}
\xi^2_{t,\delta}&=\sqrt{\frac{32 p}{3 N C}},
\end{split}\end{align}
is the best possible squeezing attainable in this region, obtained roughly at $d\geq 1$. This leads to a very broad minimum in $\xi^2_t$, depicted in Fig.~\ref{fig:Unitary}(a) for various $NC$. This trend lasts until $d\approx 1.7(CN/p)^{1/4}$, after which $\Gamma$ is no longer active, and $\xi^2_t$ now arises from the competition between the first and third terms in Eq.~(\ref{eqn:UnitSqueezing}). Further increase in $d$ worsens $\xi^2_t\sim [pd/(NC)]^{2/3}$ (independent of $\kappa$) because interactions get smaller than the spin flip rate. The optimal squeezing is thus given by Eq.~(\ref{eqn:OptimalUnitarySqueezing}).

While the region of minimum $\xi^2_{t}$ is very broad in Fig.~\ref{fig:Unitary}(a), the state area at each of these points is distinct. We show this in Fig.~\ref{fig:Unitary}(b), where $\xi^2_{t}$ vs. $A$ is plotted parametrically using $d=2\delta_*/\kappa$ as a parameter for various $NC$. The leftmost vertical sections of the curves indicate the optimal $\xi^2_{t,\delta}$, but the variation in $A$ is quite dramatic. It is preferable to work at larger values of $d$, potentially sacrificing a few dB of $\xi^2_{t}$ in exchange for a substantially smaller area, as has been pointed out before~\cite{Zhang2015,Braverman2019}.

{\it Measurement and unitary evolution:} A natural question to ask is whether a combination of measurements and unitary evolution, operating at some finite value of $d$ for a given $\eta\neq 0$, can improve upon the two limiting situations described in the previous sections. 

In Fig.~\ref{fig:Unitary}(c) we show the results of simulating numerically Eqs.~(\ref{eqn:LargeNFluc}), where at any given detuning $d$ and $\eta$, squeezing has been optimized over time within a time window of $s\in\,[0,50]$ (this accounts for the fact that at $\delta_*=0$ the optimal is only reached asymptotically). From the curves shown for different $\eta$, it can be observed that $\xi^2_{t}$ is obtained either purely through measurement at $\delta_*=0$ or in the unitary limit, where the value of $\eta$ is irrelevant. Thus, the decision to use measurements vs. OAT is determined by the comparison between Eq.~(\ref{eqn:MeasurementOptimalSqueezing}) and Eq.~(\ref{eqn:OptimalUnitarySqueezing}). They are equal when the efficiency $\eta$ has the value
\begin{equation}
    \eta_c\equiv\frac{3}{16}=0.1875.
\end{equation}
When $\eta>\eta_c$, measurements are efficient enough that operating at $\delta_*=0$ is preferable. When $\eta<\eta_c$, unitary evolution will lead to a better $\xi^2_t$.

{\it Absence of spin flips:} In cycling transitions $p$ is very close to 0 and the analysis based on Eqs.~(\ref{eqn:LargeNFluc}) is no longer applicable because the state evolves beyond the gaussian regime and gets distorted, thus introducing corrections (typically called ``finite-size" or ``curvature" effects) that limit the attainable spin squeezing. In the OAT setting ($\eta=0$), this is remedied by solving Eqs.~(\ref{eqn:LargeNFluc}) with $p=0$ and adding an extra curvature term to the minimum variance~\cite{Kitagawa-Ueda1993,Schleier-Smith2010,Chu2021}, 
\begin{equation}\label{eqn:UnitSqueezingB}
    \xi^2\approx e^{\gamma_{\text{sc}} t}\bigg(\frac{e^{\gamma_{\text{sc}} t}+\Gamma N t}{\chi^2N^2 t^2}+\frac{\chi^4N^4t^4}{6N^2}\bigg).
\end{equation}
A comparison with the analytical solution of Eq.~(\ref{eqn:EffDyn}) for $p=0$ indicates that Eq.~(\ref{eqn:UnitSqueezingB}) captures accurately the time optimized $\xi^2$~\cite{SM}. Variations of $N$ or $C$ now have different effects on $\xi^2$, whereas previously they only appeared in the combination $NC$. 

When $p=0$, the time optimized $\xi^2_{t}$ shows three distinct behaviours as a function of detuning, depicted in Fig.~\ref{fig:NoSpinFlip}(a). For $d<2.3 N^{1/3}$, collective dephasing is active, competes with the curvature term and leads to $\xi^2_{t}\approx 2N^{-2/5} d^{-4/5}$~\cite{Schleier-Smith2010,Schleier-Smith2010b} and $A=1.76 N^{1/5}/d^{3/5}$. When $2.3 N^{1/3}<d<0.4 CN^{2/3}$, the optimal squeezing arises from unitary dynamics, leading to the well-known OAT result $\xi^2_{t}= 1.04 N^{-2/3}$~\cite{Kitagawa-Ueda1993} and $A=\sqrt{1.5}$, independent of $d$. For $d>0.4 CN^{2/3}$, the exponential prefactors are the main obstruction to squeezing, and lead to $\xi^2_{t}\approx 6.8 d^2/(N^2 C^2)$ and $A\approx \sqrt{e}$. Furthermore, the existence of the OAT minimum imposes a restriction on the cooperativity: $C>6 N^{-1/3}$. Otherwise, the center region in Fig.~\ref{fig:NoSpinFlip}(a) disappears.

As $p$ is increased, the dependence of $\xi^2_{t}$ on $d$ will switch from the one shown in Fig.~\ref{fig:NoSpinFlip}(a) to the one depicted in Fig.~\ref{fig:Unitary}(a). We can estimate this value of spin flip probability, which we denote as $p_{c_1}$, by equating the exact OAT result and Eq.~(\ref{eqn:OptimalUnitarySqueezing}):
\begin{equation}
    p_{c_1}=\frac{0.1C}{N^{1/3}}.
\end{equation}
In the QND setup ($d=0$) at $p=0$, the system will approach a state with no $\hat{S}_z$ variance in a timescale $\sim (\Gamma\eta)^{-1}$, but squeezing will be limited by loss of contrast. This is shown in Fig.~\ref{fig:NoSpinFlip}(b), which is obtained by solving semi-analytically Eq.~(\ref{eqn:EffDyn})~\cite{SM} for $p=d=0$, $N=100$ and averaging $\xi^2$ over different measurement trajectories. At the optimal time, the average squeezing is
\begin{equation}\label{eqn:QNDHeisen}
    \xi^2_t\approx \frac{e}{N\eta}\bigg(1+\frac{1}{C}\bigg),
\end{equation}
calculated using the model depicted in Fig.~\ref{fig:NoSpinFlip}(b) [dashed black, see~\cite{SM} for derivation], which captures reasonably well the time development of the average $\xi^2$, though individual measurement trajectories may reach better values of $\xi^2_t$ when $C\gtrsim 1$ and $\eta\approx 1$ [see Fig~\ref{fig:NoSpinFlip}(b), shaded area]. Equating Eq.~(\ref{eqn:QNDHeisen}) and Eq.~(\ref{eqn:MeasurementOptimalSqueezing}) indicates that this minimum can be reached when $p<p_{c_2}=e^2(C+1)^2/(2NC\eta)$.

{\it Summary and conclusions:} These results form a coherent picture, summarized in terms of a few key statements, and shown schematically in Fig.~\ref{fig:NoSpinFlip}(c) and (d). 
\begin{itemize}
    \item $\underline{\eta>0.1875}$: QND is better than OAT for any value of spin flip probability $p$.
    \item $\underline{\eta<0.1875}$: OAT dominates over QND for $p$ close to 1/2. As $p$ is reduced OAT saturates to the curvature-limited ideal minimum, but QND continues to improve according to Eq.~(\ref{eqn:MeasurementOptimalSqueezing}). QND will outperform OAT when
    $p<p_{c_3}=0.54 C\eta/N^{1/3}$ (obtained by equating Eq.~(\ref{eqn:MeasurementOptimalSqueezing}) and the squeezing at the ideal OAT minimum) as long as $\eta>2.6(1+C^{-1})/N^{1/3}$ (obtained by equating Eq.~(\ref{eqn:QNDHeisen}) and the ideal OAT minimum). Otherwise OAT ourperforms QND for all $p$ (not depicted in Fig.~\ref{fig:NoSpinFlip}).
\end{itemize}
\Rev{In~\cite{SM} we discuss  how our results can indeed be used to set bounds on the amount of squeezing achievable in  experiments where our  analysis applies~\cite{Leroux2010,Schleier-Smith2010c,Cox2016,Hosten2016,Bao2020,Braverman2019}.} Our results should also apply to two tone QND schemes~\cite{Chen2011}, which are more favourable for technical reasons but don't change fundamental scalings. Future research will involve the consequences of parking the cavity closer to atomic resonance~\cite{zhang2023stochastic,Braverman2019}, comparisons with time-reversal based unitary protocols~\cite{Davis2016,Hosten2016b,Linneman2016,Colombo2022} and including more complicated unitary dynamics \Rev{(e.g. twist-and-turn, two-axis-twisting})~\cite{li2023improving,Borregaard_2017,Hu2017} using the stochastic Schr\"{o}dinger equation formalism.

\begin{acknowledgments}
{\it Acknowledgements:} During completion of this work we became aware of a related theory work examining the combination of OAT and QND in the context of Bose-Einstein condensates~\cite{Fuderer2023}. We thank C. Luo and M. Miklos for a careful reading and comments on the manuscript. This work is supported by the AFOSR grant FA9550-18-1-0319, by the DARPA and ARO grant W911NF-16-1-0576, the ARO single investigator award W911NF-19-1-0210, the NSF PHY1820885, NSF JILA-PFC PHY-1734006 and NSF QLCI-2016244 grants, and by NIST.
\end{acknowledgments}
\putbib[library]
\end{bibunit}
\balancecolsandclearpage
\onecolumngrid
\appendix
\begin{bibunit}
\setcounter{equation}{0}
\setcounter{figure}{0}
\setcounter{page}{1}
\renewcommand{\theequation}{S\arabic{equation}}
\renewcommand{\thefigure}{S\arabic{figure}}
\renewcommand{\bibnumfmt}[1]{[S#1]}
\renewcommand{\citenumfont}[1]{S#1}

\renewcommand{\braket}[1]{\big\langle#1\big\rangle}
\newcommand{\Pbig}{\bBigg@{1.25}}
\newcommand{\braketo}[1]{\big\langle#1\big\rangle}

\begin{center}
\textbf{\large Trade-offs between unitary and measurement induced spin squeezing in cavity QED: Supplementary Material}
\end{center}
In this Supplementary Material we show how to derive Eq.~(\ref{eqn:EffDyn}) starting from the fundamental atom-light interaction inside a cavity through adiabatic elimination of excited optical states and cavity degrees of freedom. We then explain how, starting from Eq.~(\ref{eqn:EffDyn}), we can obtain Eq.~(\ref{eqn:LargeNFluc}) and Eq.~(\ref{eqn:LargeNMean}) for second and first order correlators, respectively, by using the large $N$ and gaussianity assumptions. We continue by showing how these correlators are related to quantities of interest like spin squeezing and state area. We then portray how to obtain the optimal spin squeezing and state area in the setting of QND measurements, both in the presence and absence of spin flips. We finalize by repeating the same calculations for OAT dynamics. 

\section*{Effective evolution equation}
Here we show how to obtain Eq.~(\ref{eqn:EffDyn}) starting from the fundamental atom-light interaction inside a cavity. We consider $N$ three-level atoms (levels $\ket{e}$, $\ket{\downarrow}$, $\ket{\uparrow}$ with energies $\omega_e$, $\omega_{\downarrow}$ and $\omega_{\uparrow}$) in the configuration shown in Fig.~\ref{fig:Schematic}(a). The atoms interact with one cavity mode at frequency $\omega_c$, with single photon Rabi frequencies $2g_1$ and $2g_2$ for the transitions $\ket{e}\leftrightarrow\ket{\uparrow}$ and $\ket{e}\leftrightarrow\ket{\downarrow}$, respectively. The cavity has power decay linewidth $\kappa$, and the atoms spontaneously emit from $\ket{e}\to\ket{\uparrow}$ and from $\ket{e}\to\ket{\downarrow}$ with rates $\gamma_1$  and $\gamma_2$, respectively. Furthermore, the cavity is externally driven by a laser at frequency $\omega_d$ (close to $\omega_c$) and with a flux of $|\beta|^2$ photons per second. The light escaping the cavity is then collected and subjected to homodyne detection with overall efficiency $\eta$. The evolution of the combined atom-light system in the presence of all these processes is described by the following stochastic differential equation
\begin{equationS}\label{eqn:AppFullEquation}
    d\hat{\rho}&=-i\Big[\underbrace{\omega_e\hat{N}_e+\omega_\uparrow\hat{N}_{\uparrow}+\omega_\downarrow \hat{N}_{\downarrow}+\omega_c\hat{a}^\dagger\hat{a}}_{\hat{H}_{\text{free}}}+\underbrace{g_1\big(\hat{a}\hat{S}_{e\uparrow}+\text{h.c.}\big)+ g_2\big(\hat{a}\hat{S}_{e\downarrow}+\text{h.c.}\big)}_{\hat{H}_{\text{int}}}+\underbrace{\beta\sqrt{\kappa}\big(\hat{a}e^{i\omega_d t}+\text{h.c.}\big)}_{\hat{H}_{\text{drive}}},\hat{\rho}\Big]\,dt\\
&+\underbrace{\kappa\Big(\hat{a}\hat{\rho}\hat{a}^\dagger-\frac{\{\hat{a}^\dagger\hat{a},\hat{\rho}\}}{2}\Big)}_{\text{photon leakage}}\,dt+\underbrace{\gamma_1 \sum_{i=1}^N\Big(\hat{\sigma}_{\uparrow e}^i\,\hat{\rho}\,\hat{\sigma}_{e\uparrow}^i-\frac{\{\hat{\sigma}_{ee}^i,\hat{\rho}\}}{2}\Big)dt+\gamma_2 \sum_{i=1}^N\Big(\hat{\sigma}_{\downarrow e}^i\,\hat{\rho}\,\hat{\sigma}_{e\downarrow}^i-\frac{\{\hat{\sigma}_{ee}^i,\hat{\rho}\}}{2}\Big)dt}_{\text{spontaneous emission}}\\
&+\underbrace{\sqrt{\kappa\eta}\Big(\hat{a} e^{-i\phi}\hat{\rho}+\hat{\rho}\hat{a}^\dagger e^{i\phi}-\braket{\hat{a} e^{-i\phi}+\hat{a}^\dagger e^{i\phi}}\hat{\rho}\Big)}_{\text{homodyne detection}}\,dW,
\end{equationS}
where $\hat{\rho}$ is the density matrix of the atom-light system, $\hat{\sigma}_{ab}^i=\ket{a}\hspace{-0.14cm}\bra{b}_i$ ($a,b=e,\uparrow,\downarrow$) is a single particle transition operator for atom $i$, $\hat{S}_{ab}=\sum_{i=1}^N\hat{\sigma}_{ab}^i$ is a collective transition operator, $\hat{N}_{a}=\hat{S}_{aa}$ is the ``number of atoms in level $a$" operator, $\hat{a}^\dagger$ ($\hat{a}$) are bosonic creation (annihilation) operators for the cavity mode and $\phi$ is the homodyne detection angle, which is an experimentally adjustable parameter. The first line describes coherent evolution under a Hamiltonian that consists of three parts: the free uncoupled evolution of atoms and cavity, the atom-light interaction (of Jaynes-Cummings type), and the external laser drive. The second line describes the dissipative processes: photon leakage and spontaneous emission from the excited state. The third line implements homodyne detection~\cite{wiseman_milburn_2009SM,Jacobs2006SM}, which is intimately tied to photon leakage since these photons are the ones that are being measured.  

Since the cavity is detuned from the $\ket{\downarrow},\ket{\uparrow}\leftrightarrow \ket{e}$ by an amount $\pm\Delta$, the atom-light interaction becomes dispersive when $|\Delta|\gg g_{1,2}\sqrt{N},g_{1,2}\sqrt{n_{\text{phot}}}$, where $n_{\text{phot}}$ is the number of photons inside the cavity. We implement this mathematically by means of a Schrieffer-Wolff transformation. To be more precise, we define a dressed density matrix $\hat{\rho}_d=e^{\hat{W}}\,\hat{\rho}\,e^{-\hat{W}}$, where 
\begin{equation}
    \hat{W}=\frac{g_1}{\Delta}(\hat{a}^\dagger\hat{S}_{\uparrow e}-\hat{a}\hat{S}_{e\uparrow})+\frac{g_2}{\Delta}(\hat{a}\hat{S}_{e\downarrow}-\hat{a}^\dagger\hat{S}_{\downarrow e})
\end{equation}
is the Schrieffer-Wolff generator (note $e^{\hat{W}}$ is unitary).  The conditions $|\Delta|\gg g_{1,2}\sqrt{N},g_{1,2}\sqrt{n_{\text{phot}}}$ then mean that $\hat{W}$ acts perturbatively in the states of interest. The evolution equation for $\hat{\rho}_d$ can then be written down by considering the following relations:
\begin{equationS}
    e^{\hat{W}}(\hat{H}_{\text{free}}+\hat{H}_{\text{int}})e^{-\hat{W}}&=\hat{H}_{\text{free}}+\frac{g_1^2}{\Delta}\big[\hat{a}^\dagger\hat{a}(\hat{S}_{\uparrow\uparrow}-\hat{S}_{ee})-\hat{S}_{e\uparrow}\hat{S}_{\uparrow e}\big]-\frac{g_2^2}{\Delta}\big[\hat{a}^\dagger\hat{a}(\hat{S}_{\downarrow\downarrow}-\hat{S}_{ee})-\hat{S}_{e\downarrow}\hat{S}_{\downarrow e}\big]+O(g_{1,2}^3/\Delta ^2)\\[4pt]
    e^{\hat{W}}\,\hat{a}\,e^{-\hat{W}}&=\hat{a}+\frac{1}{\Delta}(g_2\hat{S}_{\downarrow e}-g_1\hat{S}_{\uparrow e})+O(g_{1,2}^2/\Delta^2)\\[4pt]
    e^{\hat{W}}\,\hat{\sigma}^i_{\uparrow e}\,e^{-\hat{W}}&=\hat{\sigma}_{\uparrow e}^i+\frac{g_1\hat{a}}{\Delta}(\hat{\sigma}_{\uparrow\uparrow}^i-\hat{\sigma}^i_{ee})-\frac{g_2\hat{a}}{\Delta}\hat{\sigma}_{\uparrow\downarrow}^i+O(g_{1,2}^2/\Delta^2)\\[4pt]
    e^{\hat{W}}\,\hat{\sigma}^i_{\downarrow e}\,e^{-\hat{W}}&=\hat{\sigma}_{\uparrow e}^i+\frac{g_2\hat{a}}{\Delta}(\hat{\sigma}_{\downarrow\downarrow}^i-\hat{\sigma}^i_{ee})+\frac{g_1\hat{a}}{\Delta}\hat{\sigma}_{\downarrow\uparrow}^i+O(g_{1,2}^2/\Delta^2)\\[4pt]
    e^{\hat{W}}\,\hat{\sigma}^i_{e e}\,e^{-\hat{W}}&=\hat{\sigma}^i_{ee}+\frac{1}{\Delta}\Big[\hat{a}(g_1\hat{\sigma}_{e\uparrow}^i-g_2\hat{\sigma}^i_{e\downarrow})+\hat{a}^\dagger(g_1\hat{\sigma}^i_{\uparrow e}-g_2\hat{\sigma}^i_{\downarrow e})\Big]\\
    &+\frac{\hat{a}^\dagger\hat{a}}{\Delta}\Big[g_1^2(\hat{\sigma}_{\uparrow\uparrow}^i-\hat{\sigma}_{ee}^i)+g_2^2(\hat{\sigma}_{\downarrow\downarrow}^i-\hat{\sigma}_{ee}^i)-g_1g_2(\hat{\sigma}_{\uparrow\downarrow}^i+\hat{\sigma}_{\downarrow\uparrow}^i)\Big]\\
    &+\frac{1}{2\Delta}\Big[g_1 g_2(\hat{\sigma}_{e\downarrow}^i\hat{S}_{\uparrow e}+\hat{\sigma}_{e\uparrow}^i\hat{S}_{\downarrow e})-g_1^2\hat{\sigma}_{e\uparrow}^i\hat{S}_{\uparrow e}-g_2^2\hat{\sigma}_{e\downarrow}^i\hat{S}_{\downarrow e}+\text{ h.c.}\Big]+O(g_{1,2}^3/\Delta^3)
\end{equationS}
We have expanded the transformed operators only to the order that will be relevant for the evolution equation of $\hat{\rho}_d$. However, before writing this equation down we will point out a few features that will allow us to simplify the result. First, single particle decoherence will now include terms of the form $\gamma_{1,2} g_{1}\hat{\sigma}^i_{e\uparrow}\hat{a}\hat{\rho}_d/\Delta,\gamma_{1,2} g_{2}\hat{\sigma}^i_{e\downarrow}\hat{a}\hat{\rho}_d/\Delta$ (and complex conjugates), which transfer atoms from the ground manifold to the excited state. However, they oscillate at frequency $\pm\Delta$ (due to $\hat{H}_{\text{free}}$) and can thus be neglected since they will effectively contribute corrections of size $\gamma^2_{1,2}g^2/\Delta^3$ to the evolution equation. These terms become relevant when $\Delta \leq \gamma$ in which case a more refined treatment is required where both spontaneous emission and $\hat{H}_{\text{free}}$ are included as the zeroth order term of a dissipative Schrieffer-Wolff transformation. We do not attempt this since it is considerably more challenging to implement from a technical standpoint and is outside our regime of interest anyway, but this means that $\Delta \gtrsim \gamma$ is another assumption of our treatment. After neglecting these terms, the resulting equation will preserve the $\ket{\uparrow},\ket{\downarrow}$ manifold. In the dressed basis, all atoms are in the ground states, and any admixture with $\ket{e}$ is taken care of by the dressing, which allows us to get rid of any term of the form $\hat{\sigma}_{\uparrow e}^i\hat{\rho},\,\hat{\sigma}_{\downarrow e}^i\hat{\rho},\,\hat{\sigma}_{e e}^i\hat{\rho}$ (and their complex conjugates). All these simplifications are generic after adiabatic elimination of $\ket{e}$. However, in the specific case we are considering, there is also a large energy difference between $\ket{\uparrow}$ and $\ket{\downarrow}$ of $\pm 2\Delta$. This means that terms of the form $\hat{\sigma}^i_{\uparrow\downarrow}\hat{\rho}_d\hat{\sigma}_{\uparrow\uparrow}^i,\,\hat{\sigma}^i_{\uparrow\downarrow}\hat{\rho}_d$ (and similar), which arise due to interfering decay pathways, oscillate very fast and can be neglected too. Taking all of this into consideration, the resulting equation for $\hat{\rho}_d$ is
\begin{equationS}\label{eqn:AppSomeEquation}
    d\hat{\rho}_d&=-i\Big[\omega_c\hat{a}^\dagger\hat{a}+\omega_{\downarrow}\hat{N}_{\downarrow}+\omega_{\uparrow}\hat{N}_{\uparrow}+\frac{(g_1^2-g_2^2)}{2\Delta}\hat{a}^\dagger\hat{a}\,N+\frac{g_1^2+g_2^2}{2\Delta}\hat{a}^\dagger\hat{a}\,(\hat{S}_{\uparrow\uparrow}-\hat{S}_{\downarrow\downarrow})+\beta\sqrt{\kappa}\big(\hat{a}e^{i\omega_d t}+\text{h.c.}\big),\hat{\rho}_d\Big]\,dt\\[4pt]
    &+\frac{\gamma_1 g_1^2}{\Delta^2}\sum_{i=1}^N\Big(\hat{\sigma}_{\uparrow\uparrow}^i\hat{a}\,\hat{\rho}_d\,\hat{a}^\dagger\hat{\sigma}_{\uparrow\uparrow}^i-\frac{\{\hat{a}^\dagger\hat{a}\,\hat{\sigma}_{\uparrow\uparrow}^i,\hat{\rho}_d\}}{2}\Big)\,dt+\frac{\gamma_2 g_2^2}{\Delta^2}\sum_{i=1}^N\Big(\hat{\sigma}_{\downarrow\downarrow}^i\hat{a}\,\hat{\rho}_d\,\hat{a}^\dagger\hat{\sigma}_{\downarrow\downarrow}^i-\frac{\{\hat{a}^\dagger\hat{a}\,\hat{\sigma}_{\downarrow\downarrow}^i,\hat{\rho}_d\}}{2}\Big)\,dt\\[4pt]
    &+\frac{\gamma_1 g_2^2}{\Delta^2}\sum_{i=1}^N\Big(\hat{\sigma}_{\uparrow\downarrow}^i\hat{a}\,\hat{\rho}_d\,\hat{a}^\dagger\hat{\sigma}_{\downarrow\uparrow}^i-\frac{\{\hat{a}^\dagger\hat{a}\,\hat{\sigma}_{\downarrow\downarrow}^i,\hat{\rho}_d\}}{2}\Big)\,dt +\frac{\gamma_2 g_1^2}{\Delta^2}\sum_{i=1}^N\Big(\hat{\sigma}_{\downarrow\uparrow}^i\hat{a}\,\hat{\rho}_d\,\hat{a}^\dagger\hat{\sigma}_{\uparrow\downarrow}^i-\frac{\{\hat{a}^\dagger\hat{a}\,\hat{\sigma}_{\uparrow\uparrow}^i,\hat{\rho}_d\}}{2}\Big)\,dt\\[4pt]
    &+\kappa\Big(\hat{a}\hat{\rho}_d\hat{a}^\dagger-\frac{\{\hat{a}^\dagger\hat{a},\hat{\rho}\}}{2}\Big)\,dt+\sqrt{\kappa\eta}\Big(\hat{a} e^{-i\phi}\hat{\rho}_d+\hat{\rho}_d\hat{a}^\dagger e^{i\phi}-\braket{\hat{a} e^{-i\phi}+\hat{a}^\dagger e^{i\phi}}\hat{\rho}_d\Big)\,dW.
\end{equationS}
Note that in the photon leakage, laser drive and homodyne detection terms, the leading order contribution is zeroth order in $g/\Delta$. The single particle decoherence terms, which are $O(g^2/\Delta^2)$, describe off-resonant scattering of cavity light (hence the presence of the operators $\hat{a}$ and $\hat{a}^\dagger$) and involve both spin flip and spin conserving processes. We also now define $\hat{S}_z=(\hat{S}_{\uparrow\uparrow}-\hat{S}_{\downarrow\downarrow})/2$ and $\hat{N}_g=\hat{S}_{\uparrow\uparrow}+\hat{S}_{\downarrow\downarrow}$ for notational simplicity. Furthermore, since we are considering only the ground manifold, we will now write $\hat{\sigma}^i_{\uparrow\uparrow}+\hat{\sigma}^i_{\downarrow\downarrow}=1$.

The drive will establish a field inside the cavity that will depend on the drive-cavity detuning. However, the frequency of the cavity now depends on the state of the atoms, through the $\hat{a}^\dagger\hat{a}\,\hat{S}_z$ term, and will create an effective nonlinear atom-atom interaction. We take this into account perturbatively by analyzing fluctuations about the cavity field that would be established at the mean field level, taking into account that the average inversion is $\braket{\hat{S}_z}=0$ for the initial conditions we want to consider [and it will remain so, as per Eq.~(\ref{eqn:AppSomeEquation})]. We thus define the fluctuation $\hat{b}=\hat{a}-\alpha e^{-i\omega_d t}$, where
\begin{equation}
    \alpha=-\frac{i\sqrt{\kappa}\beta}{-i\delta_*+\kappa/2},
\end{equation}
and $\delta_*=\omega_d-\omega_c-(g_1^2-g_2^2)N/(2\Delta)$. At this point it is convenient to move into the rotating frame of the drive. In this frame, and in terms of the fluctuation field $\hat{b}$, we approximately get
\begin{equationS}
    d\hat{\rho}_d&=-i\Big[-\delta_*\hat{b}^\dagger\hat{b}+\frac{(\omega_{\downarrow}+\omega_{\uparrow})}{2}\hat{N}_{g}+\Big(2\Delta+\frac{(g_1^2+g_2^2)|\alpha|^2}{\Delta}\Big)\hat{S}_z+\frac{(g_1^2+g_2^2)(\alpha\hat{b}^\dagger+\alpha^*\hat{b})\hat{S}_z}{\Delta},\hat{\rho}_d\Big]\,dt\\[4pt]
    &+\kappa\Big(\hat{b}\hat{\rho}_d\hat{b}^\dagger-\frac{\{\hat{b}^\dagger\hat{b},\hat{\rho}\}}{2}\Big)\,dt+\frac{\big(\gamma_1 g_1^2+\gamma_2 g_2^2\big)|\alpha|^2}{2\Delta^2}\sum_{i=1}^N\Big(\hat{\sigma}_{\uparrow\uparrow}^i\,\hat{\rho}_d\,\hat{\sigma}_{\uparrow\uparrow}^i+\hat{\sigma}_{\downarrow\downarrow}^i\,\hat{\rho}_d\,\hat{\sigma}_{\downarrow\downarrow}^i-\hat{\rho}_d\Big)\,dt\\[4pt]
    &+\frac{\gamma_1 g_2^2|\alpha|^2}{\Delta^2}\sum_{i=1}^N\Big(\hat{\sigma}_{\uparrow\downarrow}^i\,\hat{\rho}_d\,\hat{\sigma}_{\downarrow\uparrow}^i-\frac{\{\hat{\sigma}_{\downarrow\downarrow}^i,\hat{\rho}_d\}}{2}\Big)\,dt +\frac{\gamma_2 g_1^2|\alpha|^2}{\Delta^2}\sum_{i=1}^N\Big(\hat{\sigma}_{\downarrow\uparrow}^i\,\hat{\rho}_d\,\hat{\sigma}_{\uparrow\downarrow}^i-\frac{\{\hat{\sigma}_{\uparrow\uparrow}^i,\hat{\rho}_d\}}{2}\Big)\,dt\\[4pt]
    &+\sqrt{\kappa\eta}\Big(\hat{b} e^{-i\phi}\hat{\rho}_d+\hat{\rho}_d\hat{b}^\dagger e^{i\phi}-\braket{\hat{b} e^{-i\phi}+\hat{b}^\dagger e^{i\phi}}\hat{\rho}_d\Big)\,dW.
\end{equationS}
In the previous equations we have symmetrized the spin conserving single particle processes by using the relation $\hat{\sigma}_{\uparrow\uparrow}^i+\hat{\sigma}_{\downarrow\downarrow}^i=1$, kept only the leading order terms (in $|\alpha|$) in the single particle decoherence contributions, and we have neglected $(g_1^2+g_2^2)\hat{b}^\dagger\hat{b}\,\hat{S}_z/\Delta$ in the Hamiltonian part, which is a nonlinear fluctuation term. It amounts to assuming that the quantum fluctuations in $\hat{b}$ that arise because of $\hat{S}_z$ are smaller than $|\alpha|$. We now adiabatically eliminate the cavity degree of freedom. This can be done by means of a dissipative Schrieffer-Wolff transformation (since $\kappa$ and $\delta$ may be comparable now), but the fastest way to obtain the result is by doing the replacement 
\begin{equation}\label{eqn:AppLinearbtoSz}
    \hat{b}\to -\frac{i(g_1^2+g_2^2)\alpha\hat{S}_z/\Delta}{-i\delta_*+\kappa/2},
\end{equation}
which is what a mean field treatment suggests will be the steady state intracavity field, on average. This leads to
\begin{equationS}
    d\hat{\rho}_d&=-i\Big[\frac{(\omega_{\downarrow}+\omega_{\uparrow})}{2}\hat{N}_{g}+\Big(2\Delta+\frac{(g_1^2+g_2^2)|\beta|^2}{\Delta}\Big)\hat{S}_z+\Big(\frac{(g_1^2+g_2^2)|\alpha|}{\Delta}\Big)^2\frac{\delta_*\,\hat{S}_z^2}{\delta_*^2+(\kappa/2)^2},\hat{\rho}_d\Big]\,dt\\[4pt]
    &+\Big(\frac{(g_1^2+g_2^2)|\alpha|}{\Delta}\Big)^2\frac{\kappa}{\delta_*^2+(\kappa/2)^2}\Big(\hat{S}_z\hat{\rho}_d\hat{S}_z-\frac{\{\hat{S}_z^2,\hat{\rho}\}}{2}\Big)\,dt+\frac{\gamma_1 g_2^2|\alpha|^2}{\Delta^2}\sum_{i=1}^N\Big(\hat{\sigma}_{\uparrow\downarrow}^i\,\hat{\rho}_d\,\hat{\sigma}_{\downarrow\uparrow}^i-\frac{\{\hat{\sigma}_{\downarrow\downarrow}^i,\hat{\rho}_d\}}{2}\Big)\,dt \\[4pt]
     &+\frac{\gamma_2 g_1^2|\alpha|^2}{\Delta^2}\sum_{i=1}^N\Big(\hat{\sigma}_{\downarrow\uparrow}^i\,\hat{\rho}_d\,\hat{\sigma}_{\uparrow\downarrow}^i-\frac{\{\hat{\sigma}_{\uparrow\uparrow}^i,\hat{\rho}_d\}}{2}\Big)\,dt+\frac{(\gamma_1 g_1^2+\gamma_2 g_2^2)|\alpha|^2}{2\Delta^2}\sum_{i=1}^N\Big(\hat{\sigma}_{\uparrow\uparrow}^i\,\hat{\rho}_d\,\hat{\sigma}^i_{\uparrow\uparrow}+\hat{\sigma}_{\downarrow\downarrow}^i\,\hat{\rho}_d\,\hat{\sigma}^i_{\downarrow\downarrow}-\hat{\rho}_d\Big)\,dt\\[4pt]
    &+\sqrt{\Big(\frac{(g_1^2+g_2^2)|\alpha|}{\Delta}\Big)^2\frac{\kappa\eta}{\delta_*^2+(\kappa/2)^2}}\Big(\hat{S}_z\hat{\rho}_d+\hat{\rho}_d\hat{S}_z-2\braket{\hat{S}_z}\hat{\rho}_d\Big)\,dW.
\end{equationS}
At this point it is convenient to introduce the parameters
\begin{equationS}
    C&=\frac{4g_1^2}{\kappa\gamma_1}=\frac{4g_2^2}{\kappa \gamma_2}\\
    \gamma_{\text{sc}}&=\frac{\gamma_1g_1^2|\alpha|^2}{\Delta^2}+\frac{\gamma_1g_2^2|\alpha|^2}{\Delta^2}+\frac{\gamma_2g_1^2|\alpha|^2}{\Delta^2}+\frac{\gamma_2g_2^2|\alpha|^2}{\Delta^2}=\frac{(\gamma_1+\gamma_2)(g_1^2+g_2^2)|\alpha|^2}{\Delta}\\
    p&=\frac{\gamma_1g_2^2+\gamma_2g_1^2}{(\gamma_1+\gamma_2)(g_1^2+g_2^2)}=\frac{2\gamma_1 \gamma_2}{(\gamma_1+\gamma_2)^2}
\end{equationS}
The cooperativity $C$ is a property of cavity geometry. Its existence implies that $g_2^2/\gamma_2=g_1^2/\gamma_1$ and hence that the rates of the $\hat{\sigma}_{\uparrow\downarrow}^i$ and $\hat{\sigma}_{\downarrow\uparrow}^i$ processes are equal. The parameter $\gamma_{\text{sc}}$ is the total scattering rate of photons off the drive and is the sum of the rates of all single particle processes, and $p$ is the spin flip probability, which is defined as the ratio between the sum of the rates of the $\hat{\sigma}_{\uparrow\downarrow}^i$ and $\hat{\sigma}_{\downarrow\uparrow}^i$ processes and $\gamma_{\text{sc}}$. We have also made a specific choice of homodyne angle, $\phi=2\arctan(2\delta/\kappa)-\pi$, to guarantee maximum measurement strength. To simplify notation, we define
\begin{equation}\label{eqn:AppEffectiveParameters}
    \chi=\Big(\frac{(g_1^2+g_2^2)|\alpha|}{\Delta}\Big)^2\frac{\delta_*}{\delta_*^2+(\kappa/2)^2}=\frac{C\gamma_{\text{sc}}d/2}{d^2+1},\hspace{1cm} \Gamma=\Big(\frac{(g_1^2+g_2^2)|\alpha|}{\Delta}\Big)^2\frac{\kappa}{\delta_*^2+(\kappa/2)^2}=\frac{C\gamma_{\text{sc}}}{1+d^2},
\end{equation}
and we have decided to express the effective parameters $\chi$ and $\Gamma$ in terms of $C$, $\gamma_{\text{sc}}$ and $d=2\delta_*/\kappa$. Omitting single particle rotation terms, we can thus express the evolution of the system in terms of the effective equation
\begin{equationS}\label{eqn:AppEffectiveEvolutionEquation}
    d\hat{\rho}_d&=-i\Big[\chi\hat{S}_z^2,\hat{\rho}_d\Big]\,dt+\Gamma\Big(\hat{S}_z\hat{\rho}_d\hat{S}_z-\frac{\{\hat{S}_z^2,\hat{\rho}\}}{2}\Big)\,dt +\frac{\gamma_{\text{sc}}p}{2}\sum_{i=1}^N\Big(\hat{\sigma}_{\uparrow\downarrow}^i\,\hat{\rho}_d\,\hat{\sigma}^i_{\downarrow\uparrow}+\hat{\sigma}_{\downarrow\uparrow}^i\,\hat{\rho}_d\,\hat{\sigma}^i_{\uparrow\downarrow}-\hat{\rho}_d\Big)\,dt\\[4pt]
    &+\frac{\gamma_{\text{sc}}(1-p)}{2}\sum_{i=1}^N\Big(\hat{\sigma}_{\uparrow\uparrow}^i\,\hat{\rho}_d\,\hat{\sigma}^i_{\uparrow\uparrow}+\hat{\sigma}_{\downarrow\downarrow}^i\,\hat{\rho}_d\,\hat{\sigma}^i_{\downarrow\downarrow}-\hat{\rho}_d\Big)\,dt+\sqrt{\Gamma \eta}\Big(\hat{S}_z\hat{\rho}_d+\hat{\rho}_d\hat{S}_z-2\braket{\hat{S}_z}\hat{\rho}_d\Big)\,dW.
\end{equationS}
Equations~(\ref{eqn:AppEffectiveEvolutionEquation}) and~(\ref{eqn:AppEffectiveParameters}) are the same as Eq.~(\ref{eqn:EffDyn}) and Eq.~(\ref{eqn:EffParam}) in the main text once we replace $\hat{\sigma}_{\uparrow\downarrow}=\hat{\sigma}^+$, $\hat{\sigma}_{\downarrow\uparrow}=\hat{\sigma}^-$, $\hat{\sigma}_{\uparrow\uparrow}=(1+\hat{\sigma}_z)/2$ and $\hat{\sigma}_{\downarrow\downarrow}=(1-\hat{\sigma}_z)/2$

We also summarize here the assumptions that went into this derivation. First $\Delta \gg \gamma, g_{1,2}\sqrt{N}, g_{1,2}|\alpha|$ ($|\alpha|^2$ is the number of photons in the cavity) for the elimination of the the excited state. Then $|\alpha|^2 \gg \braket{\hat{b}^\dagger\hat{b}}$ to treat the resulting dispersive atom-light interaction perturbatively during adiabatic elimination of the cavity. This translates into
\begin{equation}
    \frac{(g_1^2+g_2^2)\sqrt{\braket{\hat{S}_z^2}}}{\Delta\sqrt{\delta_*^2+(\kappa/2)^2}}<\frac{2(g_1^2+g_2^2)\sqrt{\braket{\hat{S}_z^2}}}{\Delta\kappa}\ll 1\to C\sqrt{\braket{\hat{S}_z^2}}\ll \frac{\Delta}{(\gamma_1+\gamma_2)/2}.
\end{equation}
We have considered the worst case scenario with $\delta_*=0$ since this is relevant for QND measurements. For the initial conditions we are considering, $\braket{\hat{S}_z^2}\sim N$, so this assumption becomes $NC\ll \Delta\sqrt{N}/(\gamma_1+\gamma_2)$. \Rev{A violation of this condition indicates that quantum fluctuations are strong enough that the relation between the adiabatically eliminated $\hat{b}$ and $\hat{S}_z$ can no longer be captured by the simple linear relation Eq.~(\ref{eqn:AppLinearbtoSz}) and a nonlinear approximation is required~\cite{Davis2016SM}.}
\section*{Large N equations for first and second order correlators}
Here we compute the evolution equations for relevant observables. We calculate things exactly as much as possible, and apply approximations only in the end. The equations for linear observables are obtained by multiplying Eq.~(\ref{eqn:EffDyn}) with the appropriate operator and tracing the result. First we do averages of $\hat{S}_{x,y,z}$
\begin{equationS}
    d\braket{\hat{S}_x}&=\Big[-\chi\braket{\{\hat{S}_z,\hat{S}_y\}}-\frac{\Gamma}{2}\braket{\hat{S}_x}-\frac{\gamma_{\text{sc}}}{2}\braket{\hat{S}_x}\Big]\,dt+\sqrt{\Gamma\eta}\Big[\braketo{\{\hat{S}_x,\hat{S}_z\}}-2\braket{\hat{S}_x}\braket{\hat{S}_z}\Big]\,dW\\
    d\braket{\hat{S}_y}&=\Big[\chi\braket{\{\hat{S}_z,\hat{S}_x\}}-\frac{\Gamma}{2}\braket{\hat{S}_y}-\frac{\gamma_{\text{sc}}}{2}\braket{\hat{S}_y}\Big]\,dt+\sqrt{\Gamma\eta}\Big[\braketo{\{\hat{S}_y,\hat{S}_z\}}-2\braket{\hat{S}_y}\braket{\hat{S}_z}\Big]\,dW\\
    d\braket{\hat{S}_z}&=-\gamma_{\text{sc}}p\braket{\hat{S}_z}\,dt+2\sqrt{\Gamma \eta}\Big[\braket{\hat{S}_z^2}-\braket{\hat{S}_z}\braket{\hat{S}_z}\Big]\,dW,
\end{equationS}
where $\{\}$ is the anticommutator. Relevant two-point correlators are
\begin{equationS}
    d\braket{\hat{S}_z^2}&=-2\gamma_{\text{sc}}p \bigg[\braket{\hat{S}_z}^2-\frac{N}{4}\bigg]\,dt+2\sqrt{\Gamma\eta}\Big[\braket{\hat{S}_z^3}-\braket{\hat{S}_z^2}\braket{\hat{S}_z}\Big]\,dW,\\
    d\braketo{\{\hat{S}_z,\hat{S}_y\}}&=\Big[2\chi\braketo{\{\hat{S}_z,\{\hat{S}_z,\hat{S}_x\}\}}-\frac{\Gamma}{2}\braketo{\{\hat{S}_z,\hat{S}_y\}}-\gamma_{\text{sc}}(p+1/2)\braketo{\{\hat{S}_z,\hat{S}_y\}}\Big]\,dt\\
    &\hspace{5cm}+\sqrt{\Gamma\eta}\Big[\braketo{\{\hat{S}_z,\{\hat{S}_z,\hat{S}_y\}\}}-2\braketo{\{\hat{S}_z,\hat{S}_y\}}\braket{\hat{S}_z}\Big]\,dW\,\\
    d\braket{\hat{S}_y^2}&=\Big[\chi\braket{\{\hat{S}_y,\{\hat{S}_z,\hat{S}_x\}\}}+\Gamma\braket{\hat{S}_x^2}-\Gamma\braket{\hat{S}_y^2}-\gamma_{\text{sc}}\Big(\braket{\hat{S}_y^2}-N/4\Big)\Big]\,dt\\
    &\hspace{6cm}+\sqrt{\Gamma\eta}\Big[\braket{\{\hat{S}_z,\hat{S}_y^2\}}-2\braket{\hat{S}_z}\braket{\hat{S}_y^2}\Big]\,dW
\end{equationS}
We can obtain exact equations for variances and covariances, keeping in mind the product rule $d(ab)=adb+bda+da\,db$ and the Ito rules $dW^2=dt$, $dt dW=dt^2=0$ to get the right stochastic equations. Thus
\begin{equationS}
    d\,\text{Var}(\hat{S}_z)&=\Big[-4\Gamma \eta\,\text{Var}(\hat{S}_z)^2-2\gamma_{\text{sc}}p\Big(\text{Var}(\hat{S}_z)-\frac{N}{4}\Big)\Big]\,dt+2\sqrt{\Gamma\eta}\Big[\braket{\hat{S}_z^3}-3\braket{\hat{S}_z^2}\braket{\hat{S}_z}+2\braket{\hat{S}_z}\hspace{-0.05cm}^3\Big]\,dW\\
    d\,\text{Cov}&=\bigg(\frac{\chi}{2}\Big[\braket{\{\{\hat{S}_x,\hat{S}_z\},\hat{S}_z\}}-2\braket{\hat{S}_z}\braket{\{\hat{S}_x,\hat{S}_z\}}\Big]-4\Gamma\eta \text{Var}(\hat{S}_z)\,\text{Cov}-\Big[\frac{\Gamma}{2}+\gamma_{\text{sc}}(p+1/2)\Big]\text{Cov}\bigg)\,dt\\
    &\hspace{1cm}+\frac{\sqrt{\Gamma\eta}}{2}\Big[\braket{\{\hat{S}_z,\{\hat{S}_y,\hat{S}_z\}\}}-4\braket{\{\hat{S}_z,\hat{S}_y\}}\braket{\hat{S}_z}-4\braket{\hat{S}_y}\braket{\hat{S}_z^2}+8\braket{\hat{S}_y}\braket{\hat{S}_z}\hspace{-0.05cm}^2\Big]\,dW\\
    d\,\text{Var}(\hat{S}_y)&=\bigg(\chi\Big[\braket{\{\hat{S}_y,\{\hat{S}_z,\hat{S}_x\}\}}-2\braket{\hat{S}_y}\braket{\{\hat{S}_z,\hat{S}_x\}}\Big]+\Gamma\Big[\braket{\hat{S}_x^2}-\text{Var}(\hat{S}_y)\Big]-\gamma_{\text{sc}}\Big[\text{Var}(\hat{S}_y)-N/4\Big]-4\Gamma\eta\,\text{Cov}^2\bigg)\,dt\\
    &\hspace{1cm}+\sqrt{\Gamma\eta}\Big[\braket{\{\hat{S}_y^2,\hat{S}_z\}}-2\braket{\hat{S}_z}\braket{\hat{S}_y^2}-2\braket{\hat{S}_y}\braketo{\{\hat{S}_y,\hat{S}_z\}}+4\braket{\hat{S}_y}^2\braket{\hat{S}_z}\Big]\,dW,
\end{equationS}
where $\text{Var}$ and $\text{Cov}=\braket{\{\hat{S}_z,\hat{S}_y\}}/2-\braket{\hat{S}_y}\braket{\hat{S}_z}$ are  short-hands for variance and the $YZ$ covariance, respectively. We now begin with the large $N$ approximation. To apply it, we define the mean field equations of motion, obtained by replacing $(\braket{\hat{S}_x},\braket{\hat{S}_y},\braket{\hat{S}_z})\to N(X,Y,Z)/2$ and $(\braket{\hat{S}_z^2},\braket{\{\hat{S}_z,\hat{S}_y\}},\braket{\{\hat{S}_z,\hat{S}_x\}})\to N^2(Z^2,2YZ,2XZ$)/4 (factorization), where $X,Y,Z$ are functions of time. This results in
\begin{equationS}
    \dot{X}=-\chi N YZ -\frac{\gamma_{\text{sc}}}{2}X,\hspace{1cm} \dot{Y}=\chi N XZ -\frac{\gamma_{\text{sc}}}{2}Y,\hspace{1cm} \dot{Z}=-\gamma_{\text{sc}}p Z,
\end{equationS}
where we have further neglected $\Gamma\ll \Gamma N$, since it is of higher order in the $1/N$ expansion. By keeping $\gamma_{\text{sc}}$ we are implicitly assuming that $\gamma_{\text{sc}}>\Gamma$. This is not necessarily always the case, especially when the single particle cooperativity $C\sim 1$. However, in all cases $\gamma_{\text{sc}}$ is the obstruction to squeezing, and so we will always keep it. The mean field equations, together with initial conditions $X(0)=1, Y(0)=Z(0)=0$ result in $X= e^{-\gamma_{\text{sc}} t/2}, Y=Z=0$. We define the deviations from mean field $(\delta\hat{S}_x,\delta\hat{S}_y,\delta\hat{S}_z)=(\hat{S}_x-Ne^{-\gamma_{\text{sc}}t}/2,\hat{S}_y,\hat{S}_z)$. In the large $N$ limit, fluctuations and standard deviations are typically of size $\sqrt{N}$, so we further define $[v_{zz},v_{zy},v_{yy}]=4[\text{Var}(\hat{S}_z),\text{Cov},\text{Var}(\hat{S}_y)]/N$ and $(\hat{x},\hat{y},\hat{z})=2(\delta\hat{S}_x,\delta\hat{S}_y,\delta\hat{S}_z)/\sqrt{N}$. Initially, all of these quantities are $\sim 1$, so we expect that this will remain so for some time. In terms of these definitions (but still working exactly),
\begin{equationS}\label{eqn:AppMean}
    d\braket{\hat{x}}&=\Big[-\frac{\chi \sqrt{N}}{2}\braket{\{\hat{z},\hat{y}\}}-\frac{\Gamma\sqrt{N}e^{-\gamma_{\text{sc}}t/2}}{2}-\frac{\Gamma}{2}\braket{\hat{x}}-\frac{\gamma_{\text{sc}}}{2}\braket{\hat{x}}\Big]\,dt+\frac{\sqrt{\Gamma N \eta}}{2}\Big[\braket{\{\hat{x},\hat{z}\}}-2\braket{\hat{x}}\braket{\hat{z}}\Big]\\
    d\braket{\hat{y}}&=\Big[\chi N e^{-\gamma_{\text{sc}}t/2}\braket{\hat{z}}-\frac{\gamma_{\text{sc}}}{2}\braket{\hat{y}}+\frac{\chi \sqrt{N}}{2}\braket{\{\hat{z},\hat{x}\}}-\frac{\Gamma}{2}\braket{\hat{y}}\Big]\,dt+\sqrt{\Gamma N\eta}\,v_{zy}\,dW\\
    d\braket{\hat{z}}&=-\gamma_{\text{sc}}p\,\braket{\hat{z}}\,dt+\sqrt{\Gamma N \eta}\,v_{zz}\,dW
\end{equationS}
and
\begin{equationS}\label{eqn:AppFluc}
    d v_{zz}&=\Big[-\Gamma N \eta v_{zz}^2-2\gamma_{\text{sc}}p(v_{zz}-1)\Big]\,dt+\sqrt{\Gamma N \eta}\Big[\braket{\hat{z}^3}-3\braket{\hat{z}^2}\braket{\hat{z}}+2\braket{\hat{z}}^3\Big]\,dW\\
    d v_{zy}&=\bigg(\chi N e^{-\gamma_{\text{sc}}t/2}v_{zz}-\Gamma N v_{zz} v_{zy}-\gamma_{\text{sc}}(p+1/2) v_{zy}+\frac{\chi\sqrt{N}}{4}\Big[\braket{\{\{\hat{x},\hat{z}\},\hat{z}\}}-2\braket{\hat{z}}\braket{\{\hat{x},\hat{z}\}}\Big]-\frac{\Gamma}{2}v_{zy}\bigg)\,dt\\
    &\hspace{1cm}+\frac{\sqrt{\Gamma\eta N}}{4}\Big[\braketo{\{\hat{z},\{\hat{y},\hat{z}\}\}}-4\braketo{\{\hat{z},\hat{y}\}}\braket{\hat{z}}-4\braket{\hat{y}}\braket{\hat{z}^2}+8\braket{\hat{y}}\braket{\hat{z}}^2\Big]\,dW\\
    d v_{yy}&=\bigg(2\chi N e^{-\gamma_{\text{sc}}t/2}v_{zy}+\Gamma N e^{-\gamma_{\text{sc}} t}-\gamma_{\text{sc}}(v_{yy}-1)-\Gamma N\eta v_{zy}^2\\
    &+\frac{\chi\sqrt{N}}{2}\Big[\braketo{\{\hat{y},\{\hat{z},\hat{x}\}\}}-2\braket{\hat{y}}\braketo{\{\hat{z},\hat{x}\}}\Big]+2\Gamma\sqrt{N}e^{-\gamma_{\text{sc}} t/2}\braket{\hat{x}}+\Gamma\braket{\hat{x}}^2-\Gamma\braket{\hat{y}^2}+\Gamma\braket{\hat{y}}^2\bigg)\,dt\\
    &+\frac{\sqrt{\Gamma\eta N}}{2}\Big[\braketo{\{\hat{y}^2,\hat{z}\}}-2\braket{\hat{z}}\braketo{\hat{y}^2}-2\braket{\hat{y}}\braketo{\{\hat{y},\hat{z}\}}+4\braketo{\hat{y}}^2\braket{\hat{z}}\Big]\,dW
\end{equationS}
Note that $\braket{\hat{z}}$ and $\braket{\hat{y}}$ are not zero since they are fluctuations with respect to the mean field values, not with respect to the true mean values of $\hat{z},\hat{y}$. The reason for this choice is that it makes the relative $N$-dependent size of various quantities more transparent. So far Eq.~(\ref{eqn:AppMean}) and Eq.~(\ref{eqn:AppFluc}) are exact given that they are just rewritings. In the large $N$ limit we omit terms $\propto \chi\sqrt{N},\Gamma\sqrt{N},\Gamma\ll \chi N,\Gamma N$. The resulting evolution equations preserve gaussianity. Since the initial state is gaussian, the stochastic terms in the equations for $v_{zz},v_{zy},v_{yy}$ drop out. We are thus led to
\begin{equationS}\label{eqn:AppAllEqs}
    \dot{v}_{zz}&=-\Gamma N\eta v_{zz}^2-2\gamma_{\text{sc}} p(v_{zz}-1)\hspace{4.9cm} dz=-\gamma_{\text{sc}}p z dt+\sqrt{\Gamma N \eta }v_{zz}\,dW\\
    \dot{v}_{yz}&=\chi N e^{-\gamma_{\text{sc}}t/2}-\Gamma N v_{zz} v_{zy}-\gamma_{\text{sc}}(p+1/2)v_{zy}\hspace{3.125cm} dy=\Big(\chi N e^{-\gamma_{\text{sc}} t/2}z-\frac{\gamma_{\text{sc}}}{2} y\Big)\,dt+\sqrt{\Gamma N\eta }v_{zy}\, dW\\
    \dot{v}_{yy}&=2\chi N e^{-\gamma_{\text{sc}} t/2} v_{zy}+\Gamma N e^{-\gamma_{\text{sc}}t }-\Gamma N \eta v_{zy}^2-\gamma_{\text{sc}}(v_{yy}-1),
\end{equationS}
which are Eq.~(\ref{eqn:LargeNFluc}) and Eq.~(\ref{eqn:LargeNMean}) in the main text, with $z=\braket{\hat{z}}$,  $y=\braket{\hat{y}}$ and $\braket{\hat{S}_x}=\frac{N}{2} e^{-\gamma_{\text{sc}} t/2}$. At the same time, the measurement record becomes
\begin{equation}
    dq=2\sqrt{\Gamma \eta}\braket{\hat{S}_z}\,dt+dW=\sqrt{\Gamma N \eta}\,z\,dt+dW
\end{equation}
\section{Spin squeezing from second order correlators}
To leading order in $N$, the Bloch vector is pointing along $+x$ with length $N e^{-\gamma_{\text{sc}}t/2}/2$. We thus need to consider fluctuations along the $y$ and $z$ directions. To be more specific, we define the matrix
\begin{equation}
    \begin{pmatrix}
    v_{zz}&v_{zy}\\
    v_{zy}&v_{yy}
    \end{pmatrix}.
\end{equation}
The minimal and maximal variances are, respectively, the smallest and largest eigenvalues of this matrix:
\begin{equation}
    v_{\text{min}}=\frac{v_{zz}+v_{yy}}{2}-\sqrt{\frac{(v_{zz}-v_{yy})^2}{4}+v_{zy}^2},\hspace{0.5cm}v_{\text{max}}=\frac{v_{zz}+v_{yy}}{2}+\sqrt{\frac{(v_{zz}-v_{yy})^2}{4}+v_{zy}^2},
\end{equation}
and the normalization we have chosen for $v_{zz},v_{yy},v_{zy}$ ensures that a spin coherent state has $v_{\text{min}}=v_{\text{max}}=1$. The naive state area is defined as the square root of the product of minimum and maximum variances. Thus
\begin{equation}
    A^*=\sqrt{v_{\text{min}}\times v_{\text{max}}}=\sqrt{v_{zz} v_{yy}-v_{zy}^2}.
\end{equation}
The Wineland squeezing parameter and normalized state area are defined by dividing by the Bloch vector length (normalized by $N/2$) squared. Thus
\begin{equation}
    \xi^2=\frac{v_{\text{min}}}{e^{-\gamma_{\text{sc}}t}}=e^{\gamma_{\text{sc}}t}\bigg[\frac{v_{zz}+v_{yy}}{2}-\sqrt{\frac{(v_{zz}-v_{yy})^2}{4}+v_{zy}^2}\bigg],\hspace{0.5cm} A=\frac{\sqrt{v_{\text{min}}\times v_{\text{max}}}}{e^{-\gamma_{\text{sc}}t}}=e^{\gamma_{\text{sc}}t}\sqrt{v_{zz} v_{yy}-v_{zy}^2},
\end{equation}
which is Eq.~(\ref{eqn:optsqueeze}).
\section*{Optimal squeezng with QND}
\subsection{With spin flips}
The equations for $v_{zz}$ and $z$ are decoupled from the rest and can be solved exactly. We write the solution in the limit that $\Gamma N\eta \gg \gamma_{\text{sc}}p$. Then
\begin{equationS}
    \dot{v}_{zz}&=-\Gamma N\eta \Bigg(v_{zz}+\frac{\gamma_{\text{sc}}p}{\Gamma N \eta}+2\sqrt{\frac{\gamma_{\text{sc}}p}{2\Gamma N \eta}}\sqrt{1-\frac{\gamma_{\text{sc}}p}{2\Gamma N \eta}}\Bigg)\Bigg(v_{zz}+\frac{\gamma_{\text{sc}}p}{\Gamma N \eta}-2\sqrt{\frac{\gamma_{\text{sc}}p}{2\Gamma N \eta}}\sqrt{1-\frac{\gamma_{\text{sc}}p}{2\Gamma N \eta}}\Bigg)\\
    &\approx  -\Gamma N \eta\bigg(v_{zz}+\sqrt{2\frac{\gamma_{\text{sc}}p}{\Gamma N \eta}}\bigg)\bigg(v_{zz}-\sqrt{\frac{2\gamma_{\text{sc}}p}{\Gamma N \eta}}\bigg).
\end{equationS}
We want to solve $z$ in terms of the measurement record, which is the accesible information, not in terms of $dW$ so we rewrite its equation as $dz=-(\Gamma \eta N v_{zz}+\gamma_{\text{sc}} p)z\,dt+\sqrt{\Gamma N \eta}\,v_{zz}\,dq$. The solutions for $v_{zz}$ and $z$ (with $v_{zz}(0)=1$ and $z(0)=0$) are then
\begin{equation}
    v_{zz}(t)\approx \epsilon \coth\bigg(\frac{\Gamma N \eta t+1}{\Gamma N \eta\tau}\bigg),\hspace{1cm} z(t)\approx \frac{\sqrt{2\gamma_{\text{sc}}p}}{\sinh\Big[\frac{\Gamma N\eta t+1}{\Gamma N\eta \tau}\Big]}\int_0^t\cosh\bigg[\frac{\Gamma N\eta s+1}{\Gamma N\eta \tau}\bigg]\,dq(s),
\end{equation}
where $\epsilon=\sqrt{2\gamma_{\text{sc}} p/(\Gamma N\eta)}$ and $\tau^{-1}=2\sqrt{\gamma_{\text{sc}}\Gamma N \eta \gamma_{\text{sc}} p}$. Note that $\Gamma N \eta \tau=1/\epsilon\gg 1$. When $t\ll \tau$ spin flips are not active, and the solutions reduce to $v_{zz}(t)=(1+\Gamma N\eta t)^{-1}$ and $z(t)\approx \sqrt{\Gamma N\eta}(1+\Gamma N\eta t)^{-1}\int_0^t dq(s)$, which is the time-averaged measurement record, in accordance with well-known results. In the opposite limit, when $t\gtrsim \tau$ and spin flips are active, we get $v_{zz}(t)\approx \sqrt{\epsilon}(1+2e^{-2t/\tau})$ and $z(t)\approx \sqrt{2\gamma_{\text{sc}} p}\int_0^t e^{-(t-s)/\tau}\,dq(s)$. Thus, the $z$ deflection of the Bloch vector is determined by the measurement record only within a time-window of size $\tau$. Spin flips have effectively introduced a memory time.
\subsection{Without spin flips}
In this case ($p=0$), Eq.~(\ref{eqn:EffDyn}) can be solved exactly (they can also be solved exactly for nonzero $\chi$, but this is not relevant for the QND analysis). The first thing to notice is that when $p=0$ the single particle dephasing terms commute (in the superoperator sense) with the rest of the evolution equation. We thus define $\hat{\upsilon}=e^{-\mathcal{L}_{\text{deph}}t}\hat{\rho}$, where $\mathcal{L}_{\text{deph}}=\sum_i(\mathcal{L}_{\hat{\sigma}_{\uparrow\uparrow}^i}+\mathcal{L}_{\hat{\sigma}_{\downarrow\downarrow}^i})$ so that 
\begin{equation}
    \mathrm{Tr}(\hat{S}_z\hat{\rho})=\mathrm{Tr}(\hat{S}_z\hat{\upsilon}),\hspace{0.5cm} \mathrm{Tr}(\hat{S}_z^2\hat{\rho})=\mathrm{Tr}(\hat{S}_z^2\hat{\upsilon}),\hspace{0.5cm} \mathrm{Tr}(\hat{S}_x\hat{\rho})=e^{-\gamma_{\text{sc}}t/2}\mathrm{Tr}(\hat{S}_x\hat{\upsilon}),
\end{equation}
and $\hat{\upsilon}$ satisfies
\begin{equation}
    d\hat{\upsilon}=\Gamma\Big(\hat{S}_z\hat{\upsilon}\hat{S}_z-\frac{\{\hat{S}_z^2,\hat{\upsilon}\}}{2}\Big)\,dt+\sqrt{\Gamma \eta}\Big(\hat{S}_z\hat{\upsilon}+\hat{\upsilon}\hat{S}_z-2\braket{\hat{S}_z}\Big)\,dW,
\end{equation}
and the expectation value is now taken with respect to $\hat{\upsilon}$. The matrix element $\braket{m|\hat{\upsilon}|n}$ satisfies the equation

\begin{equation}
    d\braket{m|\hat{\upsilon}|n}=-\frac{\Gamma(m-n)^2}{2}\braket{m|\hat{\upsilon}|n}\,dt+\sqrt{\Gamma\eta}(m+n)\braket{m|\hat{\upsilon}|n}\,dW-2\sqrt{\Gamma\eta}\braket{\hat{S}_z}\braket{m|\hat{\upsilon}|n}\,dW,
\end{equation}
which can be integrated taking into account the Ito rule. Up to $m$ and $n$ independent prefactors, this yields 
\begin{equation}
    \braket{m|\hat{\upsilon}|n}(t)\propto e^{-\Gamma t(m-n)^2/2}\,e^{-\Gamma \eta t(n+m)^2/2}\,e^{\sqrt{\Gamma \eta}(m+n)\Rev{\left(W_t+2\sqrt{\Gamma\eta}\int_0^t \braket{\hat{S}_z}\,dt'\right)}}\Rev{e^{-2\sqrt{\Gamma\eta}\int^t_0\braket{\hat{S}_z}\,dW_{t'}}}\,\overbrace{e^{-(m^2+n^2)/N}}^{\braket{m|\hat{\upsilon}|n}(0)},
\end{equation}
where we have approximated $\braket{m|\hat{\upsilon}|n}(0)$ and $W_t=\int_0^t \,dW$ is a Brownian process. With this explicit form for the matrix elements, we can calculate the relevant expectation values
\begin{equationS}
    \mathcal{C}&=\sum_{n=-\infty}^{\infty} e^{-2\big(\frac{1}{N}+\Gamma \eta t\big)(n-n^*)^2},\hspace{1.55cm}\mathrm{Tr}(\hat{\rho}\hat{S}_x)=\frac{N e^{-\frac{\Gamma t}{2}-\gamma_{\text{sc}} t/2}}{2\mathcal{C}}\sum_{n=-\infty}^{\infty} e^{-2\big(\frac{1}{N}+\Gamma \eta t\big)(n-n^*+\frac{1}{2})^2}\\
    \mathrm{Tr}(\hat{\rho}\hat{S}_z)&=\frac{1}{\mathcal{C}}\sum_{n=-\infty}^{\infty} e^{-2\big(\frac{1}{N}+\Gamma \eta t\big)(n-n^*)^2}n,\hspace{1cm}
    \mathrm{Tr}(\hat{\rho}\hat{S}_z^2)=\frac{1}{\mathcal{C}}\sum_{n=-\infty}^{\infty} e^{-2\big(\frac{1}{N}+\Gamma \eta t\big)(n-n^*)^2}n^2
\end{equationS}
where we have extended the sums to $\pm \infty$ and
\Rev{\begin{equation}
    n^*\equiv\frac{\sqrt{\Gamma \eta}/2}{\Gamma \eta t+1/N}\left(W_t+2\sqrt{\Gamma\eta}\int_0^t\braket{\hat{S}_z}\,dt'\right)
\end{equation}
sets the average $\braket{\hat{S}_z}$ and thus satisfies the consistency requirement
\begin{equation}
   dn^*=\frac{\Gamma\eta}{\Gamma\eta t+\frac{1}{N}}\left[\frac{\sum_{n=-\infty}^{\infty} e^{-2\big(\frac{1}{N}+\Gamma \eta t\big)(n-n^*)^2}(n-n^*)}{\sum_{n=-\infty}^{\infty} e^{-2\big(\frac{1}{N}+\Gamma \eta t\big)(n-n^*)^2}}\right]\,dt+\frac{\sqrt{\Gamma\eta}}{2\left(\Gamma \eta t+\frac{1}{N}\right)}dW_t.
\end{equation}}
Average quantities are then computed by first calculating them at fixed $n^*$ and then sampling $n^*$ from the previous stochastic equation. When $\Gamma  \eta\,t\lesssim 1$, the arguments of the sums vary slowly and can be approximated by integrals, leading to
\begin{equation}\label{eqn:AppNoSpinFlipQND}
    \xi^2=\frac{e^{(\Gamma +\gamma_{\text{sc}})t}}{1+\Gamma N\eta t}\longrightarrow \Gamma t_{\text{opt}}=\frac{1}{1+\frac{\gamma_{\text{sc}}}{\Gamma}}=\frac{C}{C+1},\hspace{0.5cm} \xi^2_{\text{opt}}=\frac{e}{\eta N}\bigg(1+\frac{1}{C}\bigg).
\end{equation}
Note that $\Gamma t_{\text{opt}}<1$ always so that $\Gamma \eta t_{\text{opt}}$ satisfies $\Gamma\eta t_{\text{opt}}\lesssim 1$ quite generically, except when $\eta=1$ and $C\gg 1$. Note also that squeezing is independent of $n^*$ within this approximation. Beyond this regime, which is a situation relevant only when $C\gtrsim 1$ and $\eta\approx 1$, the result does depend on $n^*$. In this scenario, squeezing can reach $\xi^2=2/N$ on specific trajectories (when $n^*$ is even) but the average over trajectories is still captured relatively well by Eq.~(\ref{eqn:AppNoSpinFlipQND}).

\section*{Optimal spin squeezing with OAT}
When $\eta=0$, Eqs.~(\ref{eqn:AppAllEqs}) become linear. The solutions are $y=z=0$ (no stochastic terms), $v_{zz}=1$, and
\begin{equationS}
 v_{zy}&=\frac{\chi N}{\gamma_{\text{sc}} p}e^{-\gamma_{\text{sc}} t}(1-e^{-\gamma_{\text{sc}} p t})\\
 v_{yy}&=1+\Gamma N t e^{-\gamma_{\text{sc}} t}+\frac{2(\chi N)^2}{(\gamma_{\text{sc}}p)^2}e^{-\gamma_{\text{sc}} t}\big(\gamma_{\text{sc}} pt -1+e^{-\gamma_{\text{sc}} pt}\big)\\
 A^2&=e^{2\gamma_{\text{sc}} t}(v_{yy}v_{zz}-v_{zy}^2)=e^{2\gamma_{\text{sc}} t}\bigg[1+\Gamma N t e^{-\gamma_{\text{sc}}t}+\frac{2(\chi N)^2}{(\gamma_{\text{sc}}p)^2}\bigg(\gamma_{\text{sc}} pt -\frac{3}{2}+2e^{-\gamma_{\text{sc}}p t}-\frac{e^{-2\gamma_{\text{sc}} p t}}{2}\bigg)\bigg],\\
 \xi^2&=\frac{e^{\gamma_{\text{sc}} t}(v_{zz} v_{yy} -v_{zy}^2)}{\frac{1+v_{yy}}{2}+\sqrt{\Big(\frac{1-v_{yy}}{2}\Big)^2+v_{zy}^2}}
\end{equationS}
where $A$ is the state area and $\xi^2$ is the squeezing parameter.
\subsection{With spin flips}
We expect the optimal squeezing to occur for $\gamma_{\text{sc}} t\ll 1$, so we calculate things perturbatively in $\gamma_{\text{sc}}$. This leads to $v_{zy}\approx \chi N t$, $v_{yy}\approx 1+\Gamma N t +\chi^2N^2t^2\approx \chi ^2N^2t^2$ ($\chi N t\gg 1$ since we would not have any squeezing otherwise). Then
\begin{equationS}
    A^2&\approx v_{zz} v_{yy}-v_{zy}^2\approx 1+\Gamma N t+\frac{2 (\chi N)^2\gamma_{\text{sc}} t^3}{3} \\
    \xi^2 &\approx \frac{A^2}{v_{yy}}=\frac{1+\Gamma Nt}{(\chi N t)^2}+\frac{2}{3}\gamma_{\text{sc}} p t,
\end{equationS}
in agreement with the exact treatment presented in Ref.~\cite{Chu2021SM}. For smaller detunings $d=2\delta_*/\kappa$, the collective dephasing term is active and the minimum squeezing is the result of the minimization of 
\begin{equationS}
    \xi^2\approx \frac{\Gamma/\chi}{\chi N t}+\frac{2}{3}\gamma_{\text{sc}} p t\longrightarrow \xi^2_t&=\sqrt{\frac{32p}{3NC}}\sqrt{1+\frac{1}{d^2}},\hspace{0.5cm}\gamma_{\text{sc}} p\,t_{\text{opt}}=\frac{3\xi^2_t}{4},\hspace{0.5cm}A^2=2\sqrt{6}\sqrt{\frac{N C}{p d^2(d^2+1)}}
\end{equationS}
where we have used Eq.~(\ref{eqn:EffParam}). This expression for squeezing leads to Eq.~(\ref{eqn:OptimalUnitarySqueezing}) for $d \gtrsim 1$. Note also that, assuming $p$ is not too small, the optimal time $t_{\text{opt}}$ satisfies $\gamma_{\text{sc}} t_{\text{opt}} \ll 1$, as promised. Even if $\xi^2_t$ is relatively insensitive to $d$ at large values of $d$, the area $A$ gets smaller at larger detunings, which is desirable. As $d$ is further increased, $\Gamma$ gets small very quickly, and the optimal squeezing is then the result of minimizing (assuming $d\gg 1$)
\begin{equationS}
    \xi^2=\frac{1}{(\chi N t)^2}+\frac{2}{3}\gamma_{\text{sc}} p t\longrightarrow \xi^2_t=3^{1/3}\bigg(\frac{2p d}{NC}\bigg)^{2/3}, \hspace{0.5cm} \gamma_{\text{sc}} p t_{\text{opt}}=\xi^2_t, \hspace{0.5cm}A^2\approx 3
\end{equationS}
Squeezing gets worse as $d$ increases, but the area is reasonably small, so it's a good idea to operate at the crossover between these two regimes. We estimate the detuning value at which this happens by equating the expressions for squeezing on both sides. This leads to
\begin{equation}
    d_{\text{crossover}}\approx \frac{8}{3^{5/4}}\bigg[\frac{NC}{2p}\bigg]^{1/4}\approx 2\bigg[\frac{NC}{2p}\bigg]^{1/4}
\end{equation}
\subsection{No spin flips}\label{sec:AppOATNoSpinFlips}
\begin{figure}
    \centering
    \includegraphics[width=0.98\textwidth]{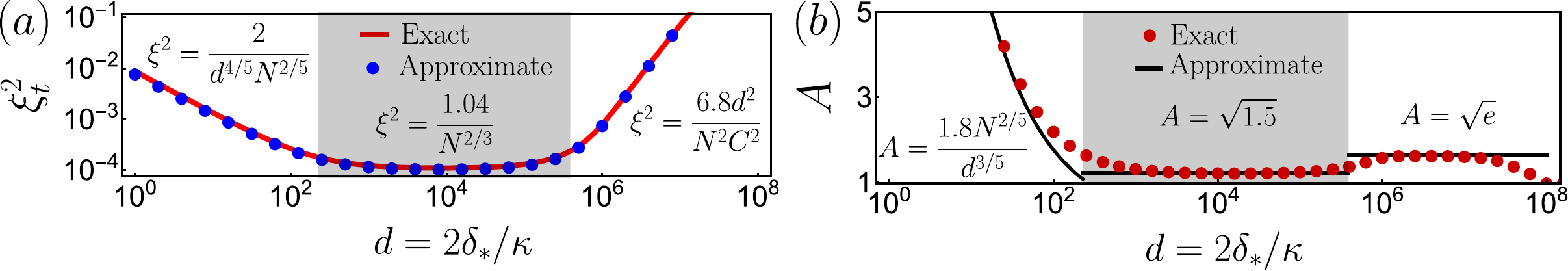}
    \caption{(a) Comparison of time optimized $\xi^2_t$ using the exact Eqs.~(\ref{eqn:AppExactOAT}) (solid red) and approximate Eq.~(\ref{eqn:AppNoSpinFlip}) (blue dots) as a function of detuning $d=2\delta_*/\kappa$ for $N=10^6$ and $C=10^2$. Boundary between different shaded regions are given by $d_{c1}=2.3N^{1/3}$ and $d_{c2}=0.4CN^{2/3}$. (b) Area at the optimal squeezing time as a function of detuning. Red dots are calculated using Eqs.~(\ref{eqn:AppExactOAT}), while black lines are the approximations for each of the regions described in section.}
    \label{fig:AppF1}
\end{figure}

We set $p=0$ here, but do not assume $\gamma_{\text{sc}} t\ll 1$ necessarily. Then $v_{yy}\approx (\chi N t)^2e^{-\gamma_{\text{sc}} t}$, $ (A^*)^2\approx v_{yy}\times v_{\text{min}}$ and
\begin{equation}
   v_{\text{min}}\approx \frac{e^{\gamma_{\text{sc}} t}+\Gamma N t}{(\chi N t)^2}+\underbrace{\frac{(\chi N t)^4}{6N^2}}_{\text{curvature}},
\end{equation}
where the extra curvature term is a correction that is not captured by Eq.~(\ref{eqn:AppAllEqs}) since it is of next to leading order in $1/N$. Furthermore, loss of contrast due to $\gamma_{\text{sc}}$ may be relevant. With $\braket{\hat{S}_x}\approx N e^{-\gamma_{\text{sc}} t}/2$, the squeezing parameter is
\begin{equation}\label{eqn:AppNoSpinFlip}
    \xi^2=e^{\gamma_{\text{sc}} t}\bigg[\frac{e^{\gamma_{\text{sc}} t}+\Gamma N t}{(\chi N t)^2}+\frac{(\chi N t)^4}{6N^2}\bigg].
\end{equation}
In fact, for $p=0$ and $\eta=0$, Eq.~(\ref{eqn:EffDyn}) can be solved exactly since all the terms in the master equation commute (in the superoperator sense). The exact values for relevant observables are
\begin{equationS}\label{eqn:AppExactOAT}
   \braket{\hat{S}_z^2}&=\frac{N}{4}\\
   \braket{\{\hat{S}_z,\hat{S}_y\}}&=\frac{N(N-1)}{2}e^{-(\gamma_{\text{sc}} +\Gamma)t/2}\cos(\chi t)^{N-2}\sin(\chi t)\\
   \braket{\hat{S}_y^2}&=\frac{N}{4}(1-e^{-\gamma_{\text{sc}}t})+\frac{N(N+1)e^{-\gamma_{\text{sc}}t}}{8}+\frac{N(1-N)e^{-\gamma_{\text{sc}}t-2\Gamma t}}{8}\cos(2\chi t)^{N-2}\\
   \braket{\hat{S}_x}&=e^{-(\gamma_{\text{sc}}+\Gamma) t/2}\cos(\chi t)^{N-1}.
\end{equationS}
A comparison of squeezing (optimized over time) obtained with the exact equations and with Eq.~(\ref{eqn:AppNoSpinFlip}) is shown in Fig.~\ref{fig:AppF1}(a) for $N=10^6$ and $C=10^2$. Since the agreement is almost perfect, we keep working with Eq.~(\ref{eqn:AppNoSpinFlip}) for analytical insight. 

As Fig.~\ref{fig:AppF1}(a) demonstrates, there can be three regimes. At smaller detunings (but still $d\gtrsim 1$), collective dephasing is active but single particle dephasing is not. The optimal squeezing is then limited by curvature and is obtained by minimizing
\begin{equation}
    \xi^2\approx \frac{\Gamma/\chi}{\chi N t}+\frac{(\chi N t)^4}{6N}\longrightarrow \xi^2_t\approx \frac{5/2}{ 3^{1/5}d^{4/5}N^{2/5}},\hspace{0.25cm}\chi N t_{\text{opt}}=\bigg(\frac{3N^2}{d}\bigg)^{1/5},\hspace{0.25cm}A\approx \sqrt{1+\frac{3.1 N^{4/5}}{d^{6/5}}}.
\end{equation}
As the detuning is increased, both the optimal squeezing and the area improve. Eventually neither collective nor single particle dephasing will be active but the optimal squeezing will still be limited by curvature and is obtained by minimizing
\begin{equation}
    \xi^2\approx \frac{1}{(\chi N t)^2}+\frac{(\chi N t)^4}{6N^2}\longrightarrow \xi^2_t\approx \frac{3^{2/3}/2}{N^{2/3}},\hspace{0.25cm} \chi N t_{\text{opt}}=(3N^2)^{1/6},\hspace{0.25cm} A\approx\sqrt{1.5}.
\end{equation}
This is the well-known OAT result. Finally, at much larger detunings $\chi$ is small enough that curvature is irrelevant, and the main limitations are now single particle dephasing and contrast decay. Hence, the optimal squeezing is obtained by minimizing
\begin{equation}
    \xi^2=\frac{e^{2\gamma_{\text{sc}}t}}{(\chi N t)^2}\longrightarrow \xi^2_t\approx \bigg(\frac{e d}{NC}\bigg)^2,\hspace{0.25cm}\gamma_{\text{sc}} t_{\text{opt}}=1,\hspace{0.25cm} A\approx \sqrt{e}.
\end{equation}
We can estimate the detunings at which the behaviour of $\xi^2_t$ changes by equating the values of the time optimized squeezing on both sides of a crossover. We find these detuning values to be
\begin{equation}
    d_{c1}=\frac{5^{5/4}}{3^{13/12}}N^{1/3}\approx 2.3 N^{1/3},\hspace{1cm} d_{c2}=\frac{3^{1/3}N^{2/3}C}{e\sqrt{2}}\approx 0.4 C N^{2/3}
\end{equation}
For the unitary region in the middle to exist, we need $d_{c1}<d_{c2}$, which requires that $C\gtrsim 6 N^{-1/3}$. Otherwise, the two outer regions in Fig. merge, and the detuning optimized squeezing will be worse than the OAT result. We show the results for the area $A$ in Fig.~\ref{fig:AppF1}(b).

\section*{Examples}
\Rev{We illustrate our results using examples that are related to the experimental system we analyzed in the main text. We will consider various QND and OAT implementations, including systems that are not directly described by Eq.~(\ref{eqn:EffDyn}).

\begin{itemize}
    \item We first analyze Ref.~\cite{Leroux2010Supp}, which implemented OAT using the same level scheme as the one in Fig.~\ref{fig:Schematic}. Using $p=1/2$, $C=0.14$ and $N=3\times 10^4$ (see second column of page 2 of Ref.~\cite{Leroux2010Supp}, where single atom cooperativity is denoted $\eta$ instead of $C$) leads to $14.5$ dB of noise reduction according to Eq.~(\ref{eqn:OptimalUnitarySqueezing}). The experiment reports $10(1)$ dB of inferred noise reduction (after subtraction of detection noise, see first column of page 4) and is further affected by fluctuations in intracavity probe power and dephasing effects that reduce the effective radius of the Bloch vector. Importantly, a theoretical analysis of this experiment was presented in Refs.~\cite{Schleier-Smith2010Supp,Schleier-Smith2010bSupp} and Eq.~(\ref{eqn:OptimalUnitarySqueezing}) is the same as the equation reported in Ref.~\cite{Schleier-Smith2010bSupp} in the presence of free space scattering.

    \item Next, we analyze Ref.~\cite{Schleier-Smith2010cSupp}, which implemented QND measurements in the same level structure as Ref.~\cite{Leroux2010Supp}. Using $p=1/2$, $\eta=0.4$ (quantum efficiency, see second column of page 2 of Ref.~\cite{Schleier-Smith2010cSupp}), and $NC=3100\times 3/2$ (see second column of page 3, the factor $3/2$ comes from discrepancies in the definition of our $C$ and the effective cooperativity in Ref.~\cite{Schleier-Smith2010cSupp}) leads to a noise reduction of $16$ dB, according to Eq.~(\ref{eqn:MeasurementOptimalSqueezing}) and $18$ dB if $\eta=1$, in agreement with the fundamental limit provided in Eq.~(20) of the Supplementary Material of Ref.~\cite{Schleier-Smith2010cSupp}. Technical noise modifies this value to $9$ dB of noise reduction and $3$ dB of metrological enhancement when including contrast decay. As discussed in the Supplementary Material of Ref.~\cite{Schleier-Smith2010cSupp}, in the ideal scenario the fundamental limit of $18$ dB is not modified by the presence of contrast decay.
    \item Now we analyze Ref.~\cite{Cox2016Supp}, which implemented QND measurements in the same level structure as Ref.~\cite{Schleier-Smith2010cSupp}, but in a configuration where the cavity is closer to resonance with $\ket{\uparrow}$ than $\ket{\downarrow}$ and with different hyperfine levels. The different choice of levels means that $p\to 0$. In this QND setup, application of Eq.~(\ref{eqn:QNDHeisen}) with $\eta=0.37$, $N=4\times 10^5$ and $C=0.044$ leads to $34$ dB of squeezing (without taking into account inhomogeneous couplings of the atoms to the cavity mode), compared to the $19$ dB of noise reduction and $17$ dB of metrological enhancement reported in Ref.~\cite{Cox2016Supp}. As discussed in the cited reference, detection noise ($25$ dB below quantum projection noise) and optomechanical effects were the main technical limitations.
    \item Then we analyze Ref.~\cite{Hosten2016Supp}, which does QND measurements in the same level structure and also the same levels as Ref.~\cite{Schleier-Smith2010cSupp}. Using $p=1/2$, $\eta=0.16$ and $NC=(5\times 10^{5})(0.78)\times 3/2$ (the factor of 3/2 arises because of discrepancies with our definition of cooperativity) leads, via Eq.~(\ref{eqn:MeasurementOptimalSqueezing}), to $25$ dB of squeezing. The cited reference reports $20$ dB of metrological enhancement and predicts $24$ dB. The difference of this prediction with our calculation originates from the broadening of the cavity linewidth due to free space scattering, a effect that we neglect and that arises as a higher order term in the adiabatic elimination of the excited state. 

    \item Let us now study the QND implementation of Ref.~\cite{Bao2020Supp}. The setup involves $N\approx 10^{11}$ atoms in a vapour cell but has no cavity. To establish a comparison to our results, we thus need to find the quantity in Ref.~\cite{Bao2020Supp} analogous to the collective cooperativity $C$ in terms of which we express all of our findings. In this case, it will be the optical depth of the atomic sample. We relate both by using the definition of $C$
\begin{equation}
    NC=\frac{4g_1^2 N}{\kappa\gamma_1},
\end{equation}
and expressing $\gamma_1$, $g_1$ and $\kappa$ in terms of atomic and cavity properties. We use the Wigner-Weisskopf expression for $\gamma_1$, Eq.~(2) of Ref.~\cite{HJKimble_1998} for $g_1$ and express $\kappa$ in terms of the cavity's free spectral range and finesse $\mathcal{F}$:
\begin{equationS}
    \gamma_1&=\frac{\omega_1^3\mu^2}{3\pi\hbar \epsilon_0 c^3}\hspace{1cm}
    g_1=\bigg(\frac{\mu^2\omega_1}{2\hbar\epsilon_0 V_m}\bigg)^{1/2}\hspace{1cm}
    \kappa=\frac{\pi c}{L\mathcal{F}},
\end{equationS}
where $\omega_1$ is the transition frequency from $\ket{\uparrow}$ to $\ket{e}$, $\mu$ is the transition dipole moment, $L$ is the cavity length and $V_m$ is the mode volume. Putting all these together we have that
\begin{equation}
    NC=\frac{6c^2}{\omega_1^2}\times \frac{N}{V_m}\times L\times \mathcal{F}.
\end{equation}
The resonant scattering cross section is $\sigma_1=4\pi c^2/\omega_1^2$, and we can take $N/V_m$ as a representative of the average atom density $n$. Then
\begin{equation}
    NC=\frac{3}{2\pi}n\sigma_1 L \mathcal{F}=\frac{3}{2\pi}(OD)\mathcal{F}
\end{equation}
where $OD=n\sigma_1 L$ is the optical depth of the sample. The experiment in Ref.~\cite{Bao2020Supp} works at an optical depth of $70$ and uses no cavity, so we set $\mathcal{F}=1$. Using these numbers to obtain a representative $NC$ and using Eq.~(\ref{eqn:MeasurementOptimalSqueezing}) with $\eta=1$ (perfect detection efficiency) and $p=1/2$ we obtain $7.5$ dB as the limit to squeezing caused by scattering from the excited state. The experiment predicts $5.3$ dB and reports $3.4$ dB of noise reduction (see last paragraph of Supplementary Information of Ref.~\cite{Bao2020Supp} for a detailed discussion), which are close to our bound, and at the level where the differences in implementation (which may introduce factors of order $1$) matter. Furthermore, at these smaller values of ``NC" contrast decay modifies the attainable metrological enhancement by amounts that are sizeable relative to the noise reduction, and Ref.~\cite{Bao2020Supp} reports $2.3$ dB of actual squeezing. 

Finally, Ref.~\cite{Bao2020Supp} uses an experimental scheme put forward in Ref.~\cite{Vasilakis2015}. The Supplementary Material of Ref.~\cite{Vasilakis2015} also includes the limitations to squeezing coming from probe-induced scattering and arrive at Eq.~(S40). Up to factors of order $1$ this is the same as Eq.~(\ref{eqn:MeasurementOptimalSqueezing}) in our paper, once the equivalence between $NC$, optical depth and finesse is made.

\item The last experiment we analyze is the OAT implementation of Ref.~\cite{Braverman2019Supp}. In this experiment, the cavity is parked close to resonance with one of the atomic transition frequencies and $p=0$, similar to Ref.~\cite{Cox2016Supp}, albeit using a different atom. For technical reasons, the experiment employs a two-tone scheme for the generation of squeezing. The first, blue-detuned, tone operates with an atom-drive detuning that falls in the center region of Fig.~\ref{fig:NoSpinFlip}(a) while the second, red-detuned, tone is in the left region of Fig.~\ref{fig:NoSpinFlip}(a) (numerical values for $\kappa$ and the atom-drive detunings are given in sections I and V of the Supplementary Material of Ref.~\cite{Braverman2019Supp}). For simplicity, we use the bound in the center region, which is the Kitagawa-Ueda limit. With $N=1000$ this leads to $20$ dB of squeezing, but inhomogeneous couplings of the atoms to the cavity mode lead to $16$ dB, as discussed in section VIII of the Supplementary Material of Ref.~\cite{Braverman2019Supp}. The experiment reports $16$ dB of noise reduction, very much in agreement with the theoretical bound, and $12$ dB of inferred metrological enhancement.
\end{itemize}

These considerations are summarized in Table~\ref{Tab:table1}.}
\begin{table}[ht]
\Rev{\begin{center} 
\begin{tabular}{ | >{\centering\arraybackslash}m{8em} |>{\centering\arraybackslash} m{12em} |>{\centering\arraybackslash}m{12.5em} |>{\centering\arraybackslash}m{10.5em}|}
  \hline
  Experiment & Inferred noise reduction & Estimation by Ref. & Our prediction\\
  \hline
  Ref.~\cite{Leroux2010Supp} (OAT) & 10 dB & -& 14.5 dB\\
  Ref.~\cite{Schleier-Smith2010cSupp} (QND)& 9 dB & 18 dB & 18 dB\\
  Ref.~\cite{Cox2016Supp} (QND)& 19 dB& - & 34 dB\\
  Ref.~\cite{Hosten2016Supp} (QND) & 20 dB & 24 dB & 25 dB\\
  Ref.~\cite{Bao2020Supp} (QND) &3.4 dB& 5.3 dB& 7.5 dB\\
  Ref.~\cite{Braverman2019Supp} (OAT) & 16 dB& 16 dB& 20 dB\\
  \hline
\end{tabular}
\vspace{1em}
\end{center}
\caption{Squeezing in experiments}\label{Tab:table1}}
\end{table}
\putbib[library2]
\end{bibunit}
\end{document}